\documentclass[11pt]{article}
\usepackage[T1]{fontenc}
\usepackage[utf8]{inputenc} 
\usepackage{lmodern}
\usepackage[margin=1in]{geometry}
\usepackage{pdflscape}
\usepackage{amsmath}
\usepackage{amssymb}
\usepackage{graphicx}
\usepackage{tikz-cd}
\usepackage{multicol}
\usepackage{comment}
\usepackage{booktabs}
\usepackage[titletoc,title]{appendix}
\usepackage{caption}
\usepackage{threeparttable}
\usepackage{longtable,array}
\usepackage{rotating}

\usepackage{tikz}
\usepackage{pgfplots}
\pgfplotsset{compat=1.18}

\usepackage[
  backend=biber,
  style=numeric,
  sorting=none,
  natbib=true,
  hyperref=true,
  maxnames=99,
  doi=true,
  url=false,
  isbn=false,
  eprint=false
]{biblatex}
\usepackage[
  colorlinks=true,
  linkcolor=blue,
  citecolor=blue,
  urlcolor=blue
]{hyperref}
\usepackage{array}
\newcommand{\DKL}{D_{\mathrm{KL}}}
\newcommand{\DJS}{D_{\mathrm{JS}}}
\newcommand{\DeltaKL}{\Delta_{\mathrm{KL}}}
\newcommand{\Atau}{A_{\tau}}
\newcommand{\Mtau}{M_{\tau}}

\title{A Note on the Kullback--Leibler Divergence in Discretized Empirical Distributions}
\author{
  Hayami Osaki\thanks{National Institute of Science and Technology Policy, Japan. 
  Email: h-osaki@nistep.go.jp}
}
\date{\today}

\begin{document}

\maketitle

\begin{abstract}
When empirical objects are represented as discrete probability distributions, within-distribution summaries such as Shannon entropy and Hill type diversity indices describe how probability mass is spread inside each object, while Kullback--Leibler (KL) divergence provides pairwise asymmetric information. This note focuses on the KL difference
\[
\Delta_{\mathrm{KL}}(p,q)
=
D_{\mathrm{KL}}(p\|q)-D_{\mathrm{KL}}(q\|p).
\]
Although \(\Delta_{\mathrm{KL}}\) can add information beyond within-distribution summaries and symmetric overlap, its sign does not, by itself, establish support inclusion, coverage, or breadth.
It is better understood as a weighted category-wise log-ratio contrast reflecting asymmetric probability-mass placement.

The point becomes clear once the definition is written out. The aim of this note is therefore to present it in a compact, example-based form, together with a descriptive bibliometric illustration based on COVID-19-related preprint-server topic distributions.
\end{abstract}

\section{Introduction}
\label{sec:introduction}

Empirical objects are often represented as probability distributions over a common set of discrete categories. Once this representation is fixed, several different questions can be asked. One may ask how broadly probability mass is spread within each object, how much two objects overlap, or whether the discrepancy between them is directionally asymmetric. These questions are related, but they are not the same. Within-distribution summaries such as Shannon entropy and Hill type diversity indices describe the spread or concentration of probability mass inside each object, symmetric quantities such as Jensen--Shannon diversity describe pairwise overlap or separation, and directional quantities describe ordered discrepancies between two distributions.

This note considers one such directional quantity based on Kullback--Leibler (KL) divergence. For two discrete probability distributions \(p=(p_1,\ldots,p_K)\) and \(q=(q_1,\ldots,q_K)\) on a common finite category set, we write
\[
D_{\mathrm{KL}}(p\|q)
:=
\sum_{k=1}^{K} p_k \log \frac{p_k}{q_k}.
\]

In empirical applications, smoothing is often applied so that the logarithmic ratios remain finite. 
We focus on the difference between the two KL directions,
\[
\Delta_{\mathrm{KL}}(p,q)
:=
D_{\mathrm{KL}}(p\|q)-D_{\mathrm{KL}}(q\|p).
\]

The interpretive issue is simple but important for reporting empirical results. A directional scalar such as \(\Delta_{\mathrm{KL}}\) has a sign, and signs are easy to translate into substantive language: one distribution may be described as broader, more concentrated, more general, more specialized, or more extended than another. Such descriptions are useful only when they are tied to quantities that support them. The sign of \(\Delta_{\mathrm{KL}}\), by itself, does not support a claim about breadth, coverage, or inclusion. It supports a statement about the asymmetric placement of probability mass across categories.
This does not mean that \(\Delta_{\mathrm{KL}}\) is unrelated to inclusion-like patterns in empirical data. Rather, such patterns need to be diagnosed rather than inferred from the KL difference alone.

The paper is positioned as a cautionary note. It does not rely on the claim that KL divergence has commonly been treated as a literal set-inclusion measure. Rather, it takes a more general concern, familiar from discussions of asymmetric distributional measures, and spells out its consequence for the specific contrast \(\Delta_{\mathrm{KL}}\). The aim is to give a compact account of how this contrast should be interpreted when it is used with discretized empirical distributions.

The structure of the paper is as follows. 
Section~\ref{sec:related_work} briefly places the present note in relation to two fields on asymmetric distributional measures and their interpretation.

Section~\ref{sec:setup_quantities} defines the quantities used in the paper and makes explicit the category-wise contribution form of \(\Delta_{\mathrm{KL}}\). Section~\ref{sec:toy} gives small toy examples showing that the contrast is not determined by common within-distribution diversity summaries, need not match a broad-versus-narrow visual impression, and can be strongly affected by low-probability regions. Section~\ref{sec:case-study} provides a descriptive illustration using COVID-19-related preprint-server topic distributions. Sections~\ref{sec:discussion} and~\ref{sec:conclusion} summarize the interpretive lesson and the intended scope of the note.

\section{Related Work and Positioning}\label{sec:related_work}

\subsection{Textbook background and scope}
\label{subsec:textbook_background}

Let
\begin{align*}
p=(p_1,\dots,p_K), \qquad q=(q_1,\dots,q_K)
\end{align*}
be two discrete probability distributions over a common finite set of \(K\) categories. The letters \(p\) and \(q\) label the two objects being compared, while
\(k=1,\dots,K\) indexes the categories. We assume
\begin{align*}
p_k \ge 0,\qquad q_k \ge 0,\qquad
\sum_{k=1}^K p_k = \sum_{k=1}^K q_k = 1.
\end{align*}
As recalled in Section~\ref{sec:introduction}, the Kullback--Leibler (KL) divergence is given by
\begin{align*}
D_{\mathrm{KL}}(p\|q)
=
\sum_{k=1}^K p_k \log \frac{p_k}{q_k},
\end{align*}
where the term with \(p_k=0\) is defined as \(0\)\footnote{This convention is consistent with the limit \(p\log p \to 0\) as \(p\to 0\).} and \(\log\) denotes the natural logarithm.  If \(p_k>0\) and \(q_k=0\) for some category \(k\), then \(D_{\mathrm{KL}}(p\|q)\) is infinite. In empirical settings, this is one reason to apply a small amount of smoothing when finite numerical values are needed.

We write
\begin{align*}
\operatorname{supp}(p)
:=
\{\,k\in\{1,\dots,K\}: p_k>0\,\}
\end{align*}
for the support set of \(p\). KL divergence is not symmetric:
\[
D_{\mathrm{KL}}(p\|q) \neq D_{\mathrm{KL}}(q\|p)
\]
in general. For example, if \(\operatorname{supp}(p) \subsetneq \operatorname{supp}(q)\), then \(D_{\mathrm{KL}}(p\|q)\) can be finite while \(D_{\mathrm{KL}}(q\|p)\) is infinite, because \(q\) assigns positive probability to categories where \(p\) is zero. Even when both directions are finite, the two values can differ substantially whenever one distribution assigns very small probability to categories that receive non-negligible mass under the other. KL divergence is therefore a comparison between full probability assignments, not a measure of inclusion between support sets.

\subsection{Computational linguistics}\label{subsec:related_cl}

Computational linguistics offers a useful point of comparison for asymmetric distributional measures. In studies of lexical entailment and hypernymy, researchers have considered whether distributional information can capture directional semantic relations, such as the relation between a more specific term and a more general one. One influential formulation is the Distributional Inclusion Hypothesis (DIH), which roughly states that the contexts of a more specific term tend to be included among the contexts of a more general term \cite{Geffet2005-fb,Kotlerman2009-zo}. This line of work motivated asymmetric distributional measures, because symmetric similarity alone cannot represent the direction of such relations.

Later studies have cautioned that such asymmetric signals should not be interpreted too directly as semantic inclusion. They can be affected by frequency, weighting choices, and by whether a method is learning a relation between two terms or merely properties of individual terms \cite{Levy2015-os,Bott2021-pt}. For the present paper, the important point is not that this literature criticized KL divergence itself, but that it provides a domain-specific example of a broader issue: asymmetric distributional quantities can invite directional semantic interpretations, and those interpretations must be checked against what the quantity actually computes.

KL divergence also appears in this broader neighbourhood. For example, \cite{Herbelot2013-vj} use KL divergence to measure semantic content by comparing a word's contextual distribution with a background distribution, and relate this measure to hyponym--hypernym ordering. This is not a use of KL divergence as a literal inclusion measure. It is better understood as a KL-based measure of departure from a reference distribution, used in a setting where specificity, generality, and hyponymy are relevant.

The present note draws on this background in a narrow sense. It does not treat the DIH literature as a direct source for \(\Delta_{\mathrm{KL}}\). Rather, it uses it as a reminder that directional distributional quantities require interpretation at the level of their actual mass-allocation structure. For \(\Delta_{\mathrm{KL}}\), this means reading the contrast as asymmetric probability-mass placement across categories, not as semantic inclusion or support inclusion.

\subsection{Ecology}\label{subsec:related_ecology}

Ecology provides another useful point of comparison, because KL-based quantities have been used directly to describe frequency-weighted patterns of resource use or interaction specialization. A clear example is the species-level specialization index \(d'\) proposed by \cite{Bluthgen2006-qv}. In that setting, a species is represented by a distribution of interaction weights across possible partners. The question is not only whether the species interacts with a given partner, but how strongly its interactions are distributed across the available partners.

The index \(d'\) uses a standardized form of KL divergence to compare the observed interaction distribution of a species with a reference distribution based on partner availability. Thus, KL divergence is not used there as a measure of set inclusion. It is used to quantify how much the observed allocation of interaction weight departs from what would be expected from availability alone. In this sense, the direction and magnitude of the quantity are tied to a comparison between probability or weight distributions, not to a literal comparison of partner sets.

Ecological studies also caution that apparent specialization or asymmetry can be affected by abundance, rarity, and sampling effort \cite{Vazquez2007-fm,Dorado2011-dw}. These cautions are not specifically warnings about KL divergence. Rather, they reinforce a broader lesson: when an asymmetric measure is applied to empirical weighted data, its interpretation should be traced back to the weights, reference distributions, and sampling structure that define the measure.

The relevance for the present note is therefore limited but direct. Ecology provides an example in which a KL-based quantity is used constructively, while its interpretation remains tied to probability-mass allocation rather than to set inclusion. The same principle applies to \(\Delta_{\mathrm{KL}}\): its sign should be read as an asymmetric contrast in mass placement across categories, not as evidence that one distribution literally contains, covers, or extends another.

\subsection{Positioning of the present study}\label{subsec:position}

The two examples above are meant only to place the present note in context. Computational linguistics and ecology both contain cases in which asymmetric distributional quantities are linked to directional terms such as inclusion, generality, specificity, or specialization. These cases also show that such quantities need to be interpreted in terms of the distributions, weights, reference baselines, and sampling assumptions that define them.

This paper applies this general caution to a specific and elementary quantity, $\Delta_{\mathrm{KL}}(p,q)$.

We do not treat the preceding literatures as direct predecessors of this contrast, nor do we claim that KL divergence has commonly been read as literal set inclusion. Our claim is narrower: when \(\Delta_{\mathrm{KL}}\) is used as a directional summary for discretized empirical distributions, its sign should be interpreted first as a contrast in probability-mass placement, rather than as an immediate statement about breadth, coverage, or inclusion.

\section{Setup and Quantities}\label{sec:setup_quantities}
\subsection{Discrete weighted representations}

Let each empirical object be represented as a nonnegative weighted distribution over \(K\) discrete categories. The only requirement is that each object be mapped to a common discrete representation that can be normalized to a probability distribution.

For a generic object, let \(n_k\) denote the count or weight assigned to category \(k\), and define the empirical distribution
\begin{align*}
\tilde p_k := \frac{n_k}{\sum_{t=1}^K n_t}.
\end{align*}
As noted in Section~\ref{subsec:textbook_background},
$D_{\mathrm{KL}}(p\|q)$ diverges when $p$ assigns positive
probability to a category where $q$ is zero. To make the computations meaningful and numerically well-defined, we apply a small amount of smoothing so that all KL divergence values remain finite, implemented as shrinkage toward the uniform background,
\begin{align*}
p_k := (1-\lambda)\tilde p_k + \frac{\lambda}{K},
\end{align*}
where \(\lambda \ll 1\) is fixed across objects. We use the uniform background as a minimally structured baseline rather than as a substantive prior assumption.

Because smoothing assigns positive probability to every category, all working distributions in the analysis have full support. Accordingly, later references to ``support-like extension'', ``coverage'', or ``inclusion'' are used informally and not meant literally in a set-theoretic sense: they refer to substantive occupancy or to probability mass that is large relative to the smoothing floor, 
or to an explicitly stated threshold
not to literal set inclusion of the smoothed supports.\footnote{The smoothing used here is equivalent to a symmetric Dirichlet prior with uniform pseudocounts. Alternative conventions exist, including asymmetric Dirichlet priors and the \(\alpha\)-skew divergence of \cite{Lee1999-nk}, which mixes the comparison distribution into the reference rather than smoothing each distribution independently. The present paper does not depend on the particular smoothing convention. What matters for the main argument is that all categories receive positive probability and that the resulting KL comparisons are interpretable. }
This leaves open the possibility of thresholded or effective-support comparisons, but those require an explicit threshold or diagnostic quantity.

When two objects are compared, we denote the resulting probability distributions by
\begin{align*}
p = (p_1, \dots,p_K), \qquad q = (q_1,\dots,q_K).
\end{align*}

\subsection{The KL difference}

We use the difference between the two KL divergences, 
already introduced in Section~\ref{sec:introduction}, as the main asymmetric comparison:
\[
  \Delta_{\mathrm{KL}}(p,q)
    = D_{\mathrm{KL}}(p\|q) - D_{\mathrm{KL}}(q\|p).
\]
When both directions are finite, this contrast can be rewritten as
\begin{align*}
\Delta_{\mathrm{KL}}(p,q)
&=
\sum_{k=1}^K p_k \log \frac{p_k}{q_k}
-
\sum_{k=1}^K q_k \log \frac{q_k}{p_k} \\
&=
\sum_{k=1}^K (p_k+q_k)\log\frac{p_k}{q_k}.
\end{align*}
This expression shows that \(\Delta_{\mathrm{KL}}\) is a weighted
log-ratio contrast over categories. Its sign and magnitude are
determined by how probability mass is placed differently across the
common category set, not directly by a set-theoretic comparison of supports.

The expression also gives a local reading: each category contributes a signed term, 
positive when \(p_k>q_k\) and negative when \(p_k<q_k\). 
This reading is most substantive when the categories already have an external meaning, for example when they correspond to predefined entities such as species, words, subject classes, or annotated topics. 
In such cases, the largest terms can indicate which labelled categories drive the departure from a reference baseline. 
By contrast, when the categories are induced mainly by a chosen representation, discretization, or smoothing scheme, the local reading should be treated mainly as diagnostic rather than substantive.

In this paper, we use \(\Delta_{\mathrm{KL}}\) alongside
within-distribution summaries and symmetric overlap. 
The point is not that it replaces these quantities, but that it records a different aspect
of the pairwise comparison.

\subsection{Summary measures}

We use several complementary within-distribution summary measures.

First, Shannon entropy is
\begin{align*}
H(p) := -\sum_{k=1}^K p_k \log p_k,
\end{align*}
and its exponential
\begin{align*}
D_1(p) := \exp\!\bigl(H(p)\bigr)
\end{align*}
is the Hill diversity \cite{Hill1973-ay} of order \(1\) and takes values in $[1, K]$. It can be read as an effective
number of categories and is relatively sensitive to lower-probability mass. $D_1$ is maximised by the uniform distribution, so a larger value indicates a ``broader'' distribution in the sense that probability mass is spread over more categories.

Second, we define
\begin{align*}
D_2(p) := \frac{1}{\sum_{k=1}^K p_k^2},
\end{align*}
the Hill diversity of order \(2\), i.e.\ the inverse Simpson index. 
It also takes values in $[1, K]$ and reflects the breadth of the high-probability part of the distribution. A smaller value indicates that high-probability mass is concentrated in a ``narrow'' few categories.

Finally, we summarize the relation between these two perspectives by
\begin{align*}
G(p) := \log\!\left(\frac{D_1(p)}{D_2(p)}\right).    
\end{align*}

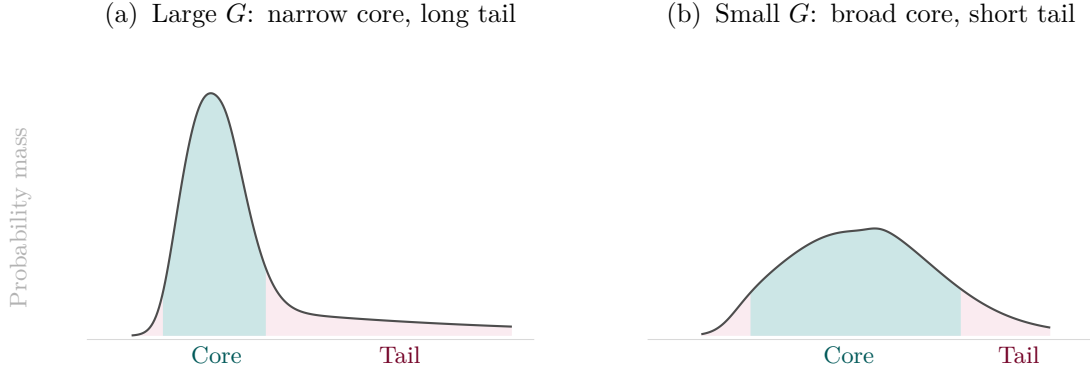
\begin{figure}[htbp]
\centering
\begin{tikzpicture}

% Panel (a): Large G --- peaked, right-skewed, long tail
\begin{axis}[
  name=panelA,
  at={(0,0)},
  width=7.5cm, height=5.5cm,
  domain=-1:8, samples=250,
  axis lines=none,
  xmin=-1.5, xmax=8.5,
  ymin=-0.08, ymax=0.72,
  clip=false,
  title style={at={(0.5,1.05)}, anchor=south, font=\small},
  title={(a)\enspace Large $G$:\enspace narrow core, long tail},
]

  % Tail fill (full distribution, light pink)
  \addplot[fill=purple!8, draw=none, domain=-0.5:8]
    {0.65 * exp(-0.5*((x-1.2)/0.7)^2) * 1/(1+exp(-5*(x-0.2)))
     + 0.08 * exp(-0.18*(x-1.5)) * 1/(1+exp(-8*(x-1.5)))} \closedcycle;

  % Core fill (narrow, teal)
  \addplot[fill=teal!20, draw=none, domain=0.2:2.5]
    {0.65 * exp(-0.5*((x-1.2)/0.7)^2) * 1/(1+exp(-5*(x-0.2)))
     + 0.08 * exp(-0.18*(x-1.5)) * 1/(1+exp(-8*(x-1.5)))} \closedcycle;

  % Curve
  \addplot[black!70, thick, domain=-0.5:8]
    {0.65 * exp(-0.5*((x-1.2)/0.7)^2) * 1/(1+exp(-5*(x-0.2)))
     + 0.08 * exp(-0.18*(x-1.5)) * 1/(1+exp(-8*(x-1.5)))};

  % Baseline
  \draw[gray!30] (-1.5,-0.01) -- (8.5,-0.01);

  % Soft labels
  \node[teal!70!black, font=\footnotesize] at (1.4, -0.05) {Core};
  \node[purple!60!black, font=\footnotesize] at (5.5, -0.05) {Tail};

\end{axis}

% Vertical axis label
\node[rotate=90, anchor=south, font=\footnotesize, text=gray!60]
  at ([xshift=-6mm]panelA.west) {Probability mass};

% Panel (b): Small G --- broad, mildly skewed, short tail
\begin{axis}[
  name=panelB,
  at={(panelA.east)}, anchor=west,
  xshift=1.5cm,
  width=7.5cm, height=5.5cm,
  domain=-1:8, samples=250,
  axis lines=none,
  xmin=-1.5, xmax=8.5,
  ymin=-0.08, ymax=0.72,
  clip=false,
  title style={at={(0.5,1.05)}, anchor=south, font=\small},
  title={(b)\enspace Small $G$:\enspace broad core, short tail},
]

  % Tail fill (full, light pink)
  \addplot[fill=purple!8, draw=none, domain=-0.3:7.5]
    {0.28 * exp(-0.5*((x-3.0)/1.8)^2) * 1/(1+exp(-4*(x-0.3)))
     + 0.04 * exp(-0.35*(x-3.5)) * 1/(1+exp(-6*(x-3.5)))} \closedcycle;

  % Core fill (broad, teal)
  \addplot[fill=teal!20, draw=none, domain=0.8:5.5]
    {0.28 * exp(-0.5*((x-3.0)/1.8)^2) * 1/(1+exp(-4*(x-0.3)))
     + 0.04 * exp(-0.35*(x-3.5)) * 1/(1+exp(-6*(x-3.5)))} \closedcycle;

  % Curve
  \addplot[black!70, thick, domain=-0.3:7.5]
    {0.28 * exp(-0.5*((x-3.0)/1.8)^2) * 1/(1+exp(-4*(x-0.3)))
     + 0.04 * exp(-0.35*(x-3.5)) * 1/(1+exp(-6*(x-3.5)))};

  % Baseline
  \draw[gray!30] (-1.5,-0.01) -- (8.5,-0.01);

  % Soft labels
  \node[teal!70!black, font=\footnotesize] at (3.0, -0.05) {Core};
  \node[purple!60!black, font=\footnotesize] at (6.8, -0.05) {Tail};

\end{axis}

\end{tikzpicture}
\caption{Schematic illustration of the tail--core gap $G = \log(D_1/D_2)$.
Both panels show a hypothetical category distribution.
In panel~(a), probability mass is concentrated in a narrow core
while a long tail extends over many low-probability categories,
producing a large ratio $D_1/D_2$ and hence a large~$G$.
In panel~(b), most of the mass is spread broadly across the core
and the tail is short, so $D_1$ and $D_2$ are close in value
and $G$ is small.
The shaded regions are suggestive rather than exact:
they indicate the part of the distribution that contributes
most to $D_2$ (core, darker) versus the additional
spread captured by $D_1$ (tail, lighter).}
\label{fig:tail-core-gap}
\end{figure}

Now, \(D_1(p)\ge D_2(p)\),\footnote{This can be shown by a straightforward argument based on Jensen's inequality.} so \(G(p)\ge 0\). A small value of \(G(p)\) means that \(D_1(p)\) and \(D_2(p)\) are close in ratio, whereas a larger value means that the ratio \(D_1(p)/D_2(p)\) is larger. In that descriptive sense, \(G\) records how far the overall effective breadth \(D_1\) exceeds the more core-weighted breadth \(D_2\). We define $G$ here as a convenient summary of the gap between $D_1$
and $D_2$; no novelty claim is attached to the quantity itself.
Intuitively, it may be read as a ``tail--core gap.''
(see Figure~\ref{fig:tail-core-gap}).

These summaries are used to distinguish overall effective breadth, high-probability breadth, and the relation between them. Their role is not to replace the KL difference, but to make clear what common within-distribution summaries do and do not capture\footnote{The choice of \(D_1\) and \(D_2\) is not unique. Other Hill orders, Rao--Stirling diversity, or domain-specific indices may be more appropriate depending on the application. The key point for the present argument is simply that breadth and concentration should not be collapsed into a single within-distribution summary when interpreting pairwise comparisons.}.

\subsection{Symmetric overlap}

To compare two objects symmetrically, we use the normalized Jensen--Shannon divergence,
\begin{align*}
D_{\mathrm{JS}}(p,q)
:=
\frac{1}{2\log 2}
\sum_{k=1}^K
\left(
p_k \log \frac{2p_k}{p_k+q_k}
+
q_k \log \frac{2q_k}{p_k+q_k}
\right).    
\end{align*}

This quantity is symmetric, takes values in \([0,1]\), and summarizes overall distributional overlap. A value of \(D_{\mathrm{JS}}=0\) indicates identical distributions, while \(D_{\mathrm{JS}}=1\) indicates distributions with no shared probability mass. In some applications, one instead uses the Jensen--Shannon distance, i.e.\ \(\sqrt{D_{\mathrm{JS}}}\), or a corresponding similarity, i.e.\ \(1-\sqrt{D_{\mathrm{JS}}}\). Nothing in the present argument depends on that choice, since only the induced symmetric ordering is relevant here.

\subsection{What these quantities do and do not determine}

The quantities introduced above play different roles. The summaries
\(D_1\), \(D_2\), and \(G\) describe within-distribution shape, while
\(D_{\mathrm{JS}}\) characterizes pairwise symmetric overlap.
The contrast \(\Delta_{\mathrm{KL}}\) describes pairwise directional
asymmetry.

Taken together, these quantities are useful because they separate
different aspects of distributional comparison: breadth within each
object, overlap between two objects, and directional mass-placement
contrast. They should not be collapsed into a single narrative. In
particular, within-distribution breadth and symmetric overlap do not in
general determine the sign or magnitude of \(\Delta_{\mathrm{KL}}\).
Conversely, when an inclusion-like interpretation is desired, it should be supported by an explicitly defined effective-support or low-probability-mass diagnostic.

The next section illustrates this point with small examples. These
examples show that \(\Delta_{\mathrm{KL}}\) can differ from common
breadth-based impressions and that its value can be strongly affected by
how probability mass is placed in low-probability regions.

\section{Toy Examples}
\label{sec:toy}

This section uses ten-category toy examples to clarify three points.
First, \(\Delta_{\mathrm{KL}}\) can provide pairwise asymmetric information not determined by common summary measures and symmetric overlap.
Second, the sign of \(\Delta_{\mathrm{KL}}\) need not agree with a simple coverage-like or broad-versus-narrow reading of category spread.
Third, the examples make visible a mass-placement pattern that is especially important for KL-based contrasts: probability mass assigned to categories that are very small under the comparison distribution can strongly affect the directional difference.
The examples are not meant to show that inclusion-like readings are always wrong, but to show that they are not entailed by \(\Delta_{\mathrm{KL}}\) alone.

Because \(D_{\mathrm{KL}}\) is unbounded, the numerical values reported below are not intended as universal benchmarks.
Their role is comparative within each example.
The relevant question is whether \(\Delta_{\mathrm{KL}}(p,q)\) reflects a directional mismatch in the pair, not whether a particular numerical value should be regarded as large in an absolute sense.

Throughout this section, qualitative terms such as ``dominant'', ``secondary'', or ``low-probability'' are purely comparative descriptions of the displayed values themselves.
They do not introduce semantic interpretations of the categories, nor do they introduce any additional threshold beyond the numerical magnitudes shown in each example.

\subsection{The KL difference adds asymmetric information beyond common summaries}
\label{subsec:asym info}

We begin with an example in which common summary measures do not determine pairwise asymmetry.
Consider
\begin{align*}
p &=
(0.320,\,0.180,\,0.140,\,0.110,\,0.090,\,0.070,\,0.040,\,0.030,\,0.015,\,0.005),\\
q &=
(0.030,\,0.180,\,0.320,\,0.110,\,0.070,\,0.090,\,0.140,\,0.015,\,0.040,\,0.005).
\end{align*}

\begin{figure}[hbt]
\centering
\begin{tikzpicture}
\begin{axis}[
    ybar,
    bar width=6.5pt,
    width=0.86\linewidth,
    height=0.42\linewidth,
    ymin=0,
    ymax=0.65,
    xlabel={Category},
    ylabel={Probability mass},
    symbolic x coords={1,2,3,4,5,6,7,8,9,10},
    xtick=data,
    enlarge x limits=0.06,
    legend style={
        at={(0.98,0.98)},
        anchor=north east,
        draw=none,
        fill=white,
        fill opacity=0.85,
        text opacity=1
    },
    legend cell align=left,
    tick align=outside,
    axis x line*=bottom,
    axis y line*=left,
]
\addplot+[
    ybar,
    fill=black!25,
    draw=black,
    area legend
] coordinates {
    (1,0.320)
    (2,0.180)
    (3,0.140)
    (4,0.110)
    (5,0.090)
    (6,0.070)
    (7,0.040)
    (8,0.030)
    (9,0.015)
    (10,0.005)
};
\addlegendentry{\(p\)}

\addplot+[
    ybar,
    fill=black!60,
    draw=black,
    area legend
] coordinates {
    (1,0.030)
    (2,0.180)
    (3,0.320)
    (4,0.110)
    (5,0.070)
    (6,0.090)
    (7,0.140)
    (8,0.015)
    (9,0.040)
    (10,0.005)
};
\addlegendentry{\(q\)}
\end{axis}
\end{tikzpicture}
\caption{
Toy Example~\ref{subsec:asym info}.
The two distributions differ only by a permutation of probability masses, so their within-distribution summaries \(D_1\), \(D_2\), and \(G\) coincide. In contrast, \(\Delta_{\mathrm{KL}}(p,q)\) has a nonzero value.
}
\label{fig:toy-example-4-1}
\end{figure}
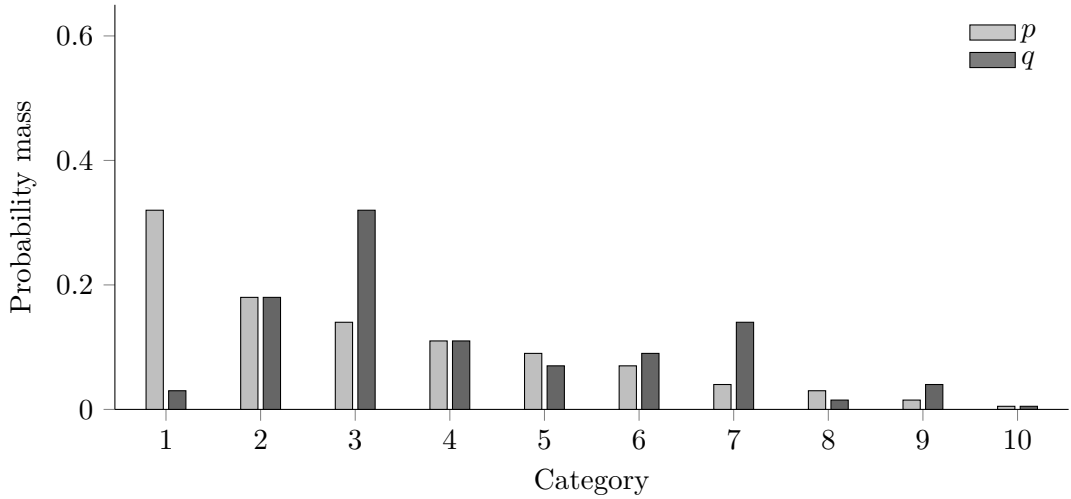

These two distributions are permutations of the same multiset of probability masses and therefore have the same values of \(D_1\), \(D_2\), and \(G\):
\begin{align*}
D_1(p)=D_1(q),\qquad
D_2(p)=D_2(q),\qquad
G(p)=G(q).
\end{align*}
They also yield only a single symmetric comparison under \(\DJS(p,q) \approx 0.156\).
Nevertheless,
\begin{align*}
\DKL(p\|q) \approx 0.603,\qquad
\DKL(q\|p) \approx 0.403,
\end{align*}
so
\begin{align*}
\DeltaKL(p,q) \approx 0.200 \neq 0.
\end{align*}

The difference arises from how probability mass is arranged across categories.
Although the two distributions agree on the multiset of probability values, they differ in where those values are placed.
This pairwise arrangement is not described by common within-distribution summaries, and a symmetric overlap measure does not distinguish the two directions.
The KL difference records this directional arrangement.
Thus, \(\DeltaKL\) can add pairwise asymmetric information beyond within-distribution shape summaries and symmetric overlap.

\subsection{The sign need not match a coverage-like impression}
\label{subsec:secondary}

We next show that the asymmetric signal in the KL difference need not align with a simple coverage-like reading.
Let \(\varepsilon=10^{-6}\), and consider
\begin{align*}
p &=
(0.075,\,0.050,\,\varepsilon,\,0.050,\,0.050,\,0.050,\,
0.575-\varepsilon,\,0.050,\,0.050,\,0.050),\\
q &=
(0.025,\,0.025,\,0.250,\,0.025,\,0.025,\,0.025,\,
0.550,\,0.025,\,0.025,\,0.025).
\end{align*}

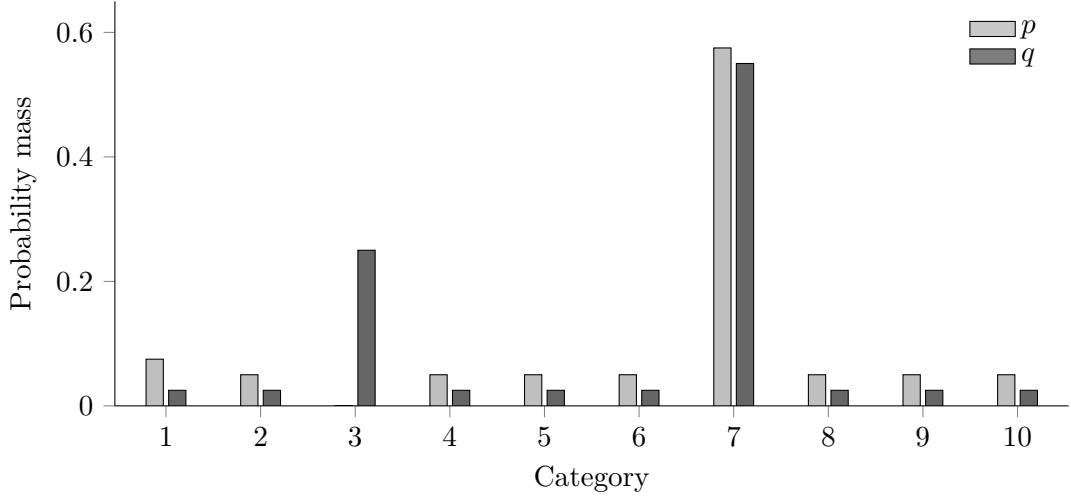
\begin{figure}[h]
\centering
\begin{tikzpicture}
\begin{axis}[
    ybar,
    bar width=6.5pt,
    width=0.86\linewidth,
    height=0.42\linewidth,
    ymin=0,
    ymax=0.65,
    xlabel={Category},
    ylabel={Probability mass},
    symbolic x coords={1,2,3,4,5,6,7,8,9,10},
    xtick=data,
    enlarge x limits=0.06,
    legend style={
        at={(0.98,0.98)},
        anchor=north east,
        draw=none,
        fill=white,
        fill opacity=0.85,
        text opacity=1
    },
    legend cell align=left,
    tick align=outside,
    axis x line*=bottom,
    axis y line*=left,
]
\addplot+[
    ybar,
    fill=black!25,
    draw=black,
    area legend
] coordinates {
    (1,0.075)
    (2,0.050)
    (3,0.000001)
    (4,0.050)
    (5,0.050)
    (6,0.050)
    (7,0.574999)
    (8,0.050)
    (9,0.050)
    (10,0.050)
};
\addlegendentry{\(p\)}

\addplot+[
    ybar,
    fill=black!60,
    draw=black,
    area legend
] coordinates {
    (1,0.025)
    (2,0.025)
    (3,0.250)
    (4,0.025)
    (5,0.025)
    (6,0.025)
    (7,0.550)
    (8,0.025)
    (9,0.025)
    (10,0.025)
};
\addlegendentry{\(q\)}
\end{axis}
\end{tikzpicture}
\caption{
Toy Example~\ref{subsec:secondary}.
Except for the third category, \(p\) assigns more probability mass than \(q\) to every category, including the main peak at category 7.
Nevertheless, \(\Delta_{\mathrm{KL}}(p,q)<0\).
The figure illustrates that a coverage-like visual impression does not determine the sign of the KL difference.
}
\label{fig:toy-example-4-2}
\end{figure}

As Figure~\ref{fig:toy-example-4-2} shows, \(p\) assigns more mass than \(q\) to every category except the third.
It also has the larger main peak:
\begin{align*}
p_7=0.575-\varepsilon>q_7=0.550.
\end{align*}
This might suggest, informally, that \(p\) more ``broadly covers'' the category set, while \(q\) differs mainly through one local bump.

This coverage-like impression is also consistent with the within-distribution summaries:
\begin{align*}
D_1(p)\approx 4.764,\qquad D_1(q)\approx 4.109,
\end{align*}
\begin{align*}
D_2(p)\approx 2.827,\qquad D_2(q)\approx 2.703,
\end{align*}
and
\begin{align*}
G(p)\approx 0.522,\qquad G(q)\approx 0.419.
\end{align*}

Taken together, \(D_1(p)>D_1(q)\), \(D_2(p)>D_2(q)\), and \(G(p)>G(q)\) may suggest, at an intuitive level, that \(p\) has both ``a broader core'' and ``a broader tail'' than \(q\).

Nevertheless,
\begin{align*}
\DKL(p\|q)\approx 0.351,\qquad
\DKL(q\|p)\approx 2.934,
\end{align*}
so
\begin{align*}
\DeltaKL(p,q)\approx -2.584.
\end{align*}

This illustrates that the KL contrast does not simply track the intuitive impression that `` \(p\) has broader support-like structure than \(q\) ''.

The next subsection isolates the mass-placement pattern responsible for this reversal more directly.

\subsection{Low-probability mass placement can strongly affect the KL difference}
\label{subsec:low prob}

The previous example showed that a coverage-like reading need not match the sign of \(\DeltaKL\), but it did not separate this effect from the overall visual impression of the distributions.
We now isolate the relevant mass-placement pattern more directly.
Consider
\begin{align*}
p &=
(0.690,\,0.100,\,0.140,\,0.010,\,0.010,\,0.010,\,0.010,\,0.010,\,0.010,\,0.010),\\
q &=
(0.690,\,0.100,\,0.001,\,0.149,\,0.010,\,0.010,\,0.010,\,0.010,\,0.010,\,0.010).
\end{align*}

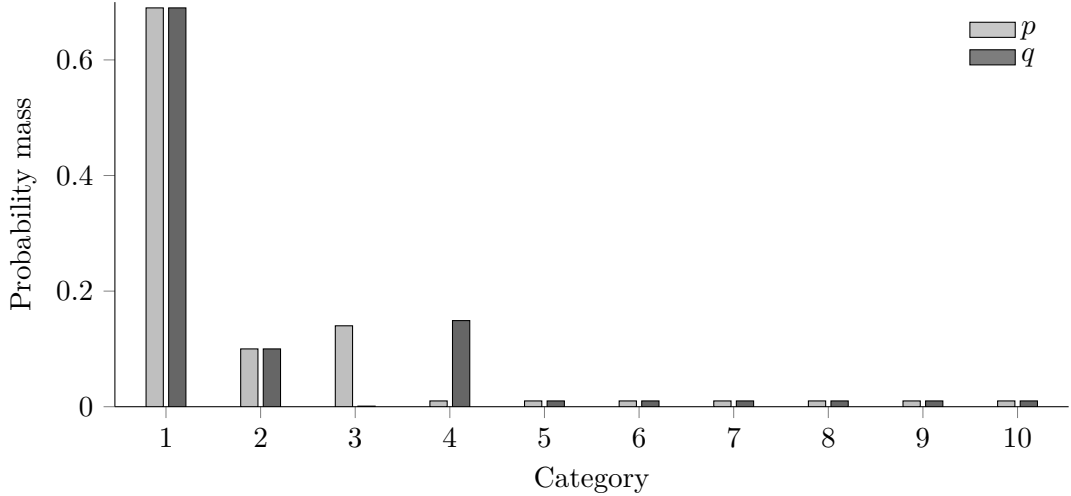
\begin{figure}[h]
\centering
\begin{tikzpicture}
\begin{axis}[
    ybar,
    bar width=6.5pt,
    width=0.86\linewidth,
    height=0.42\linewidth,
    ymin=0,
    ymax=0.70,
    xlabel={Category},
    ylabel={Probability mass},
    symbolic x coords={1,2,3,4,5,6,7,8,9,10},
    xtick=data,
    enlarge x limits=0.06,
    legend style={
        at={(0.98,0.98)},
        anchor=north east,
        draw=none,
        fill=white,
        fill opacity=0.85,
        text opacity=1
    },
    legend cell align=left,
    tick align=outside,
    axis x line*=bottom,
    axis y line*=left,
]
\addplot+[
    ybar,
    fill=black!25,
    draw=black,
    area legend
] coordinates {
    (1,0.690)
    (2,0.100)
    (3,0.140)
    (4,0.010)
    (5,0.010)
    (6,0.010)
    (7,0.010)
    (8,0.010)
    (9,0.010)
    (10,0.010)
};
\addlegendentry{\(p\)}

\addplot+[
    ybar,
    fill=black!60,
    draw=black,
    area legend
] coordinates {
    (1,0.690)
    (2,0.100)
    (3,0.001)
    (4,0.149)
    (5,0.010)
    (6,0.010)
    (7,0.010)
    (8,0.010)
    (9,0.010)
    (10,0.010)
};
\addlegendentry{\(q\)}
\end{axis}
\end{tikzpicture}
\caption{
Toy Example~\ref{subsec:low prob}.
Although the two distributions differ only in the placement of the third and
fourth probability masses, the KL is sensitive to this local
difference.
}
\label{fig:toy-example-4-3}
\end{figure}

The two distributions agree in their first two categories and in categories 5 through 10.
They differ only in categories 3 and 4.
Nevertheless,
\begin{align*}
\DKL(p\|q) \approx 0.665,\qquad
\DKL(q\|p) \approx 0.398,
\end{align*}
so
\begin{align*}
\DeltaKL(p,q)\approx 0.267>0.
\end{align*}

This example makes explicit a pattern that was already present in Section~\ref{subsec:secondary}.
Once the leading categories are held fixed, the sign and magnitude of \(\DeltaKL\) can be substantially affected by how the remaining probability mass is placed.
In particular, a small number of categories can have a large influence on the directional comparison when one distribution assigns very small probability where the other assigns substantially more.
This mass-placement pattern, rather than a vague broad-versus-narrow distinction, explains the asymmetry in this example.

\subsection{From visible toy patterns to empirical summaries}
\label{subsec:atau}

In the toy examples above, the relevant mass-placement pattern is visible by direct inspection.
In Section~\ref{subsec:secondary}, the mismatch between a coverage-like reading and the sign of \(\DeltaKL\) is visible from the displayed distribution.
In Section~\ref{subsec:low prob}, the pattern is isolated more directly because the two distributions differ only in two categories.

When the categories themselves have stable domain meanings, category-level
inspection may provide additional substantive information.
For example, if the categories correspond to species, words, subject classes,
or annotated topics, one may examine which categories contribute to the
directional difference and consider whether those differences have a meaningful
domain interpretation.
This does not mean that the KL contrast should be identified with inclusion without further diagnostics.
It only reflects the fact that KL-based quantities are sensitive to where
probability mass is placed.

In more general empirical settings, however, the categories need not have such
direct meanings.
They may be induced by preprocessing, representation learning, discretization,
clustering, or other analytical choices.
In that case, inspecting individual category indices may still identify the
numerical source of a mismatch, but it need not provide a substantive
interpretation of the categories themselves.
The case study below considers this latter situation.
It uses an embedding-based discretization, where category-level inspection is
numerically possible but not directly interpretable as inspection of
pre-defined topical classes.

\section{Case Study: COVID-19 Preprint Servers}
\label{sec:case-study}

This section illustrates, in one applied bibliometric setting, how the quantities discussed in Sections~\ref{sec:setup_quantities} and~\ref{sec:toy} can be reported together. The aim is not to provide an independent empirical proof of the preceding toy examples, but to show how the same distinction between within-object breadth, symmetric overlap, and asymmetric mass placement can be used in an applied comparison.

More specifically, we examine whether, in this dataset, the observed values of \(\Delta_{\mathrm{KL}}\) are more consistent with asymmetric low-probability mass placement than with a simple broad-versus-narrow contrast.

\subsection{Data and representation}
\label{subsec:data}

We illustrate the quantities discussed above using topic distributions
derived from COVID-19-related preprints.
The dataset used in this case study is the same as that used in
\cite{oai:nistep.repo.nii.ac.jp:02000286}.
The present analysis builds on that dataset but addresses a more specific
question: how KL differences behave when the empirical distributions
are constructed from an embedding-based discretization.
    
The dataset covers COVID-19-related preprints posted up to February 2025
and contains 47,570 records in total.
It consists of six analytical groups:
five preprint servers
(arXiv, bioRxiv, ChemRxiv, medRxiv, and SSRN)
plus SSRN Lancet, a subset extracted from SSRN.

Each preprint is represented by a text embedding
derived from its title and abstract.
To obtain a common discrete representation, we quantize the embedding space into \(K\) categories using \texttt{k-means++} clustering with a fixed random seed and then represent each server by its empirical distribution over these categories.
After smoothing with strength \(\lambda\), each server is described by a discrete probability distribution over the same common set of categories.

We report the main results for \(K = 500\) and \(\lambda = 10^{-3}\). 
We use these as a baseline specification rather than as uniquely privileged values. Robustness checks for the smoothing strength, the number of categories, and sample-size standardisation via rarefaction\footnote{Rarefaction here refers to repeated subsampling
of each server's data to the size of the smallest server, so that
all servers are compared at a common sample size. This controls for
the possibility that differences in sample size drive the observed
patterns.} are reported in Appendix~\ref{app:robustness}; the qualitative patterns described below are broadly similar across these variations, except under parameter settings that substantially change the effective representation. For a qualitative discussion of the parameter settings, also refer to Appendix~B of \cite{oai:nistep.repo.nii.ac.jp:02000286}.

\subsection{Within-server breadth profiles}
\label{subsec:breadth}

Table~\ref{tab:within_server_diversity} reports the within-server values of \(D_1\), \(D_2\), and \(G\). These three summaries capture different aspects of
distributional shape and do not in general induce a single ranking. 
In what follows, references to effective ``breadth'' and ``high-probability breadth'' are shorthand for \(D_1\) and \(D_2\),
respectively; a smaller \(D_2\) indicates a narrower concentration
of high-probability mass. \(G\) is referred to as the
``tail--core gap.''

\begin{table}[ht]
\centering
\caption{Within-server breadth summaries for \(K=500\), \(\lambda=10^{-3}\) without rarefaction.
\(D_1\): effective breadth;
\(D_2\): high-probability breadth;
\(G=\log(D_1/D_2)\): tail--core gap.}
\label{tab:within_server_diversity}
\small
\renewcommand{\arraystretch}{1.2}
\begin{tabular}{@{}lrrrr@{}}
\toprule
Server & \(n\) & \(D_1\) & \(D_2\) & \(G\) \\
\midrule
SSRN         & 3,663  & 322.2 & 220.7 & 0.378 \\
medRxiv      & 21,899 & 312.0 & 271.4 & 0.139 \\
arXiv        & 8,774  & 165.4 & 126.0 & 0.272 \\
bioRxiv      & 7,867  & 143.8 & 114.0 & 0.232 \\
SSRN Lancet  & 3,569  & 109.2 &  58.9 & 0.617 \\
ChemRxiv     & 1,756  &  59.6 &  41.5 & 0.362 \\
\bottomrule
\end{tabular}
\end{table}

The \(D_1\) and \(D_2\) rankings do not coincide. SSRN has the largest reported \(D_1\) (322.2), whereas medRxiv has the largest reported \(D_2\) (271.4). Thus, under the present representation, SSRN ranks first when lower-probability categories are weighted more heavily, whereas medRxiv ranks first when emphasis is placed on high-probability categories.

ChemRxiv has the smallest values of both \(D_1\) and \(D_2\), so it ranks lowest under both summaries.

The quantity \(G\) also varies substantially across servers. medRxiv has the smallest \(G\) (0.139), indicating that its reported \(D_1\) and \(D_2\) are relatively close. By contrast, SSRN Lancet has the largest \(G\) (0.617) despite having the second-smallest \(D_1\). This indicates not that SSRN Lancet is ``broad'' in an unrestricted sense, but that its \(D_1\) substantially exceeds its \(D_2\): \(D_2 = 58.9\) is by far the smallest reported value, whereas \(D_1 = 109.2\) is nearly twice as large.

These differences matter because no single within-server summary furnishes a pairwise directional ordering.

\subsection{Symmetric overlap and the strength of directional interpretation}

Before interpreting \(\DeltaKL\), it is useful to examine how much common structure each pair of servers actually shares. For this purpose, we use the normalised Jensen--Shannon divergence \(\DJS\) introduced in Section~\ref{sec:setup_quantities}.

Across the 15 server pairs, the normalised \(\DJS\) values range from \(\DJS \approx 0.209\) (medRxiv--SSRN) to \(\DJS \approx 0.961\) (ChemRxiv--SSRN Lancet). This wide range suggests that not all pairs should be interpreted in the same way when considering directional asymmetry. At the upper end, pairs such as ChemRxiv--medRxiv \((\DJS \approx 0.908)\) and ChemRxiv--SSRN Lancet \((\DJS \approx 0.961)\) show limited overlap in the discretised representation. At the lower end, medRxiv--SSRN \((\DJS \approx 0.209)\) exhibits much greater overlap. Detailed pairwise values are reported in Appendix~\ref{app:baseline-reference}.

This distinction matters because a nonzero value of \(\DeltaKL\) is easier to interpret when the two distributions retain substantial overlap. When overlap is very low, asymmetric comparison remains numerically well defined after smoothing, but we treat such pairs mainly as cautionary examples rather than as strong evidence for a substantive directional ordering. Section~\ref{subsec:two_pairs} therefore begins with a weak-overlap pair and then contrasts it with the highest-overlap pair in the dataset.

\subsection{\(\Atau\) as a diagnostic of asymmetric low-probability mass placement}

Section~\ref{subsec:atau} motivated a descriptive diagnostic for settings in which direct category-level interpretation is limited.
This motivation is specific rather than general.
When categories have substantive labels, inspecting the categories that drive a directional mismatch may be the most informative analysis.
In the present case study, however, the categories are cells of a vector-quantized embedding representation.
A coordinate-level inspection could identify which cells contribute to a KL difference, but the cell indices themselves do not have direct topical meanings.
We therefore use a simple threshold-based diagnostic to summarize asymmetric low-probability mass placement.

For a threshold \(\tau>0\), define
\begin{align*}
\Mtau(p\to q):=\sum_{k:\,q_k\le \tau}p_k,
\end{align*}
and
\begin{align*}
\Atau(p,q):=\Mtau(p\to q)-\Mtau(q\to p).
\end{align*}

Here \(\Mtau(p\to q)\) is the total mass that \(p\) assigns to categories that are low-probability under \(q\).
Thus, \(\Atau(p,q)>0\) indicates that \(p\) places more mass in low-probability regions of \(q\) than \(q\) places in low-probability regions of \(p\), while \(\Atau(p,q)<0\) indicates the reverse.

This quantity is used as a descriptive diagnostic of thresholded low-probability mass asymmetry. It is not introduced as a new general-purpose asymmetric index or as a literal support-inclusion test. Rather, in the empirical representation used here, it summarizes how much probability mass each distribution assigns to categories that are effectively near-absent under the other. In this limited sense, \(\Atau\) provides an effective-support diagnostic against which the behaviour of \(\DeltaKL\) can be compared.

In the present case study, we set
\begin{align*}
\tau = \frac{1}{n_{\min}} = \frac{1}{1756} \approx 5.7 \times 10^{-4},
\end{align*}
where \(n_{\min}\) is the sample size of the smallest server, ChemRxiv. This threshold corresponds approximately to categories with expected count at most one under the smallest sample size in the dataset. The rationale is pragmatic: it identifies categories that are effectively near-absent at the coarsest evidence scale present in the data.

This choice is not uniquely determined, and alternative thresholds would yield quantitatively different values of \(\Atau\). What matters for the present argument is whether, at a reasonable threshold, \(\Atau\) captures a pattern that aligns with the behaviour of \(\DeltaKL\) across server pairs. 

At the baseline specification, the smoothing floor is
\begin{align*}
\frac{\lambda}{K} = \frac{10^{-3}}{500} = 2\times 10^{-6},
\end{align*}
whereas
\begin{align*}
\tau \approx 5.7\times 10^{-4}.
\end{align*}
The diagnostic therefore targets categories well above the smoothing floor, reducing the risk that it simply reflects smoothing artefacts. Appendix~\ref{app:tau-sensitivity} reports the corresponding sensitivity checks for alternative values of \(\tau\). 

In what follows, \(A_\tau\) is used only as a diagnostic comparison, not as a criterion by which \(\Delta_{\mathrm{KL}}\) is judged. Appendix~\ref{app:tau-sensitivity} shows that \(A_\tau\) itself is stable over a moderate range of thresholds around \(1/n_{\min}\), becomes less stable at \(2/n_{\min}\), and degrades substantially only when \(\tau\) is made much more permissive.

\subsection{A cautionary weak-overlap pair and a higher-overlap pair}
\label{subsec:two_pairs}

We now examine two illustrative server pairs. The purpose is not to catalogue every pair exhaustively, but to show the two interpretive regimes most important for the argument: first, a weak-overlap pair in which a nonzero \(\DeltaKL\) should be read cautiously; second, the highest-overlap pair in the dataset, in which \(\DeltaKL\) and \(\Atau\) admit a comparatively clearer joint reading. Across all 15 server pairs, \(\DeltaKL\) ranges from \(-2.44\) to \(2.04\) and \(\Atau\) ranges from \(-0.450\) to \(0.248\). The remaining pairwise values are summarised in Appendix~\ref{app:baseline-reference}.

\paragraph{Pair 1: ChemRxiv versus medRxiv.}
This pair has very low symmetric overlap, with normalised \(\DJS \approx 0.908\). In that sense, the two topic distributions are close to separated in the discretised representation. The low-probability asymmetry diagnostic is small, with
\begin{align*}
\Atau \approx 0.016,
\end{align*}
because both distributions place large amounts of mass in categories that are low-probability for the other:
\begin{align*}
\Mtau(\text{ChemRxiv} \to \text{medRxiv}) \approx 0.912, \qquad
\Mtau(\text{medRxiv} \to \text{ChemRxiv}) \approx 0.896.
\end{align*}
Thus, although the two distributions are far apart, the asymmetry in their low-probability mass placement is slight.

At the baseline specification, the KL difference is
\begin{align*}
\DeltaKL \approx 0.828.
\end{align*}
Taken on its own, this nonzero value might invite over-interpretation. However, once the pair is viewed jointly through \(\DJS\) and \(\Atau\), it is better read as arising in a weak-overlap comparison than as evidence of a clear substantive directional ordering. The pair is therefore used here mainly as a cautionary example: a sizeable \(\DeltaKL\) can occur even when the two distributions show limited overlap and the low-probability asymmetry diagnostic is small.

\paragraph{Pair 2: medRxiv versus SSRN.}
This pair has the smallest normalised Jensen--Shannon divergence in the dataset,
\begin{align*}
\DJS \approx 0.209,
\end{align*}
and therefore provides a comparatively clearer setting for examining directional asymmetry.
The KL difference is
\begin{align*}
\DeltaKL \approx -0.445,
\end{align*}
and the low-probability asymmetry diagnostic is
\begin{align*}
A_\tau \approx -0.126.
\end{align*}
The negative value of \(A_\tau\) means that SSRN places more mass than medRxiv in categories that are low-probability for medRxiv. The negative value of \(\DeltaKL\) is consistent with this direction of asymmetry, although it reflects the full weighted log-ratio contrast rather than this thresholded diagnostic alone.

The within-server profile in Table~\ref{tab:within_server_diversity} is broadly compatible with this pairwise reading: relative to medRxiv, SSRN combines a larger \(D_1\) and a larger tail--core gap \(G\), whereas medRxiv retains the larger \(D_2\). This compatibility, however, is incidental rather than systematic. Section~\ref{sec:toy} showed that within-server summaries do not in general determine pairwise asymmetric directionality, and the alignment seen here should not be read as evidence that they do. What \(\Delta_{\mathrm{KL}}\) adds is pairwise directional information, and its agreement with \(A_\tau\) here suggests that the asymmetry is associated with SSRN placing more probability mass in regions that are low-probability for medRxiv. 

In this pair, therefore, a containment-like reading is comparatively well supported, provided that it is understood in this thresholded effective-support sense rather than as literal set inclusion. SSRN is not simply “broader” than medRxiv in every respect: medRxiv has the larger high-probability breadth \(D_2\). The pairwise result instead suggests that SSRN covers much of the shared region while placing additional mass in categories that are low-probability for medRxiv.

\subsection{Overall assessment}
\label{subsec:overall}

The toy examples provide the controlled illustration of the paper's interpretive point. The present subsection asks whether the case study shows a compatible empirical pattern.

If \(\Delta_{\mathrm{KL}}\) were primarily tracking a simple breadth contrast, then one would expect it to align with a within-object breadth comparator. In the present setup, we use \(\Delta G\) for that purpose because it is the most permissive comparator of that kind: unlike \(\Delta D_1\) or \(\Delta D_2\) alone, it already records the gap between overall and core breadth within each object. Across all 15 pairs at the baseline specification, the Spearman rank correlations between \(\Delta_{\mathrm{KL}}\) and the three within-object breadth comparators are
\begin{align*}
\rho(\Delta_{\mathrm{KL}}, \Delta G) \approx 0.35,\quad
\rho(\Delta_{\mathrm{KL}}, \Delta D_1) \approx 0.31,\quad
\rho(\Delta_{\mathrm{KL}}, \Delta D_2) \approx 0.20.
\end{align*}
\(\Delta G\) therefore yields the strongest such association among these comparators, and the analysis below uses it as the most permissive within-object breadth comparator. If even \(\Delta G\) aligns only weakly with \(\Delta_{\mathrm{KL}}\), then a simple breadth-based reading is not strongly supported in this dataset. If, by contrast, the observed ordering of \(\Delta_{\mathrm{KL}}\) is closer to asymmetric low-probability mass placement, it should align more closely with \(A_\tau\).

The corresponding Spearman rank correlation between \(\Delta_{\mathrm{KL}}\) and \(A_\tau\) is approximately
\begin{align*}
\rho(\Delta_{\mathrm{KL}}, A_\tau) \approx 0.81.
\end{align*}
Appendix~\ref{app:tau-sensitivity} shows that this contrast is stable over a moderate range of thresholds around \(\tau = 1/n_{\min}\), but weakens once \(\tau\) is made substantially larger.

Because both \(\Delta_{\mathrm{KL}}\) and \(A_\tau\) are computed from the same pair of distributions, and because \(A_\tau\) depends on an analyst-chosen threshold, this comparison should not be read as an external validation of \(\Delta_{\mathrm{KL}}\). It is a diagnostic comparison internal to the representation. Nevertheless, the result is informative: in this dataset, the ordering induced by \(\Delta_{\mathrm{KL}}\) is much closer to the thresholded low-probability mass diagnostic \(A_\tau\) than to the within-object breadth contrast \(\Delta G\). Thus, the observed KL differences are not well described as simple broad-versus-narrow contrasts. They are better described as pairwise directional contrasts in probability-mass placement, with thresholded effective-support asymmetry playing an important empirical role.

This finding does not turn \(\Delta_{\mathrm{KL}}\) into a measure of inclusion. Rather, it shows why inclusion-like interpretations must be diagnosed rather than assumed. In this case study, such a diagnostic supports a containment-like reading for some pairs, most clearly medRxiv--SSRN, while also identifying weak-overlap cases such as ChemRxiv--medRxiv where a nonzero \(\Delta_{\mathrm{KL}}\) should be read cautiously. The practical lesson is therefore not that \(\Delta_{\mathrm{KL}}\) is unrelated to inclusion-like structure, but that any such reading should be grounded in companion quantities such as \(\DJS\) and \(A_\tau\).

Robustness checks under alternative specifications of \(K\), \(\lambda\), and sample-size standardisation are reported in Appendix~\ref{app:robustness}, while sensitivity to the threshold choice in \(A_\tau\) is reported in Appendix~\ref{app:tau-sensitivity}.

\section{Discussion}
\label{sec:discussion}

The sensitivity of KL divergence to low-probability mass placement follows directly from its definition, and related cautions about asymmetric distributional measures have appeared in specific application domains, as discussed in Section 2. The purpose of the present note has been to organize this point for the directional contrast \(\Delta_{\mathrm{KL}}\) on general discrete probability distributions. The toy examples and the case study play complementary roles: the former make the relevant mass-placement patterns visible in controlled settings, while the latter illustrates how the same distinction can be used when reporting an empirical comparison.

The toy examples show why \(\Delta_{\mathrm{KL}}\) should not be reduced to a simple breadth, coverage, or inclusion-like reading. Identical within-distribution summaries do not determine \(\Delta_{\mathrm{KL}}\); a coverage-like visual impression need not determine its sign; and mass placed in categories that are very small under the comparison distribution can strongly affect the pairwise contrast. These examples do not identify a new property of KL divergence. Rather, they make explicit, in small discrete distributions, how the directional difference depends on probability-mass placement.

The case study in Section 5 has a narrower role. It does not establish a general empirical law about preprint servers or topic distributions. Instead, it shows how the same interpretive distinction can be used in one applied bibliometric setting. In that dataset, the ordering induced by \(\Delta_{\mathrm{KL}}\) is closer to the low-probability mass diagnostic \(A_\tau\) than to the within-object breadth contrast \(\Delta G\). This pattern is diagnostic rather than conclusive, but it is informative: in this case study, a substantial part of the observed KL-directional signal is empirically aligned with thresholded effective-support asymmetry.

This result should not be read as turning \(\Delta_{\mathrm{KL}}\) into a measure of inclusion. Rather, it shows the more limited point that inclusion-like interpretations can be empirically plausible when they are supported by companion diagnostics. The caution is therefore not that such interpretations are always wrong. The caution is that they should not be inferred from \(\Delta_{\mathrm{KL}}\) alone. A nonzero KL difference first indicates an asymmetric log-ratio contrast in probability-mass placement; whether that contrast is also well described as effective coverage, extension, or inclusion is an additional empirical question.

This also explains why one should not simply discard \(\Delta_{\mathrm{KL}}\) in favour of a thresholded effective-support or coverage diagnostic. If the substantive question is only whether one distribution covers the effective support of another at a chosen threshold, then a quantity such as \(A_\tau\), or a related coverage diagnostic, is the more direct object to report. But \(\Delta_{\mathrm{KL}}\) answers a different question. It summarizes a directional, category-wise log-ratio contrast over the full probability distributions, without first reducing the comparison to a binary thresholded support relation. For this reason, it can record pairwise directional information not captured by within-object summaries, symmetric overlap, or a single effective-support cutoff. The appropriate response is therefore not to replace \(\Delta_{\mathrm{KL}}\) with an inclusion score, but to report what each quantity is being used to establish.

A practical implication follows from this. Labels such as “broader”, “more concentrated”, or “more tail-heavy” are more defensible when anchored to explicitly named within-distribution quantities such as \(D_1\), \(D_2\), or \(G\). Labels such as “more inclusive”, “more covering”, or “more extended” require a different kind of support, such as an explicitly defined effective-support or low-probability-mass diagnostic. Standing alone, such labels are too coarse as glosses on a nonzero \(\Delta_{\mathrm{KL}}\). The point is not that they must never be used, but that they should be grounded in quantities that actually support them.

Several limitations should be noted. The empirical analysis uses one discretization pipeline and one family of text representations. Different embeddings, codebooks, or domains may produce different quantitative patterns. The diagnostic \(A_\tau\) depends on a threshold chosen on pragmatic grounds, and the case study uses only six analytical groups, yielding 15 pairwise comparisons. These limitations are acceptable for the illustrative role of the case study, but they should prevent the empirical results from being read as general quantitative benchmarks for preprint servers, topic distributions, or KL-based contrasts more broadly.

\section{Conclusion}
\label{sec:conclusion}

This paper examined how to interpret the KL difference
\begin{align*}
\Delta_{\mathrm{KL}}(p,q)
=
D_{\mathrm{KL}}(p\|q)-D_{\mathrm{KL}}(q\|p).
\end{align*}
when empirical objects are represented as discretized weighted distributions. The central point is interpretive. In this setting, \(\Delta_{\mathrm{KL}}\) should not be read directly as a literal measure of support inclusion. After smoothing, all working distributions have full support, and even before smoothing the quantity itself responds to probability assignments rather than to set relations. What \(\Delta_{\mathrm{KL}}\) provides is directional information about asymmetric probability-mass placement across categories.

Sections 3 and 4 made this point explicit in small discrete settings. The setup section showed that \(\Delta_{\mathrm{KL}}\) can be written as a weighted category-wise log-ratio contrast. The toy examples then illustrated that matched within-distribution summaries do not determine \(\Delta_{\mathrm{KL}}\), that a coverage-like visual impression or broad-versus-narrow contrast need not determine its sign, and that probability mass placed in categories that are very small under the comparison distribution can strongly affect the KL difference. These examples are best read as checks on interpretation rather than as a new property of KL divergence.

Section 5 provided one empirical illustration using COVID-19 preprint-server topic distributions. In that dataset, the ordering of \(\Delta_{\mathrm{KL}}\) across server pairs is closer to the low-probability mass diagnostic \(A_\tau\) than to the within-object contrast \(\Delta G\). This pattern is only diagnostic, given the small number of pairs and the threshold dependence of \(A_\tau\), but it is consistent with the view that the KL difference is a directional mass-placement contrast and that, in some empirical settings, this contrast may be strongly aligned with thresholded effective-support asymmetry.

The resulting guidance is limited but useful for reporting empirical comparisons. \(\Delta_{\mathrm{KL}}\) need not be avoided. It can record pairwise directional information not captured by within-distribution summaries or symmetric overlap alone. It records a full-distribution directional contrast that is not reducible to a single thresholded support relation. However, its sign should not be glossed directly as inclusion, breadth, or generality. Such substantive labels require support from explicitly defined companion quantities and from the meaning of the representation itself. For discretized weighted distributions, \(\Delta_{\mathrm{KL}}\) is best reported as a directional contrast sensitive to asymmetric probability-mass placement. If an inclusion-like interpretation is intended, it should be supported separately, for example by a thresholded effective-support diagnostic such as \(A_\tau\), rather than inferred from the KL difference alone.

%%%%%%%%%
\section*{Acknowledgements}
This article is based in part on a joint report with Yuko Ito, Kazuhiro Hayashi and Hitoshi Koshiba, published as \cite{oai:nistep.repo.nii.ac.jp:02000286}.

The author used generative AI tools for limited assistance in idea development, code drafting, and interpretive support during the preparation of this manuscript. All substantive judgments, verification of the results, and final decisions regarding the content were made by the author, who takes full responsibility for the manuscript.
%%%%
\clearpage

\printbibliography

@ARTICLE{Hill1973-ay,
  title        = {Diversity and evenness: A unifying notation and its
                  consequences},
  author       = {Hill, M O},
  journaltitle = {Ecology},
  publisher    = {Wiley},
  volume       = {54},
  issue        = {2},
  pages        = {427--432},
  date         = {1973-03-01},
  urldate      = {2026-05-19},
  language     = {en},
  doi          = {10.2307/1934352}
}

@ARTICLE{Vazquez2007-fm,
  title        = {Species abundance and asymmetric interaction strength in
                  ecological networks},
  author       = {Vázquez, Diego P and Melián, Carlos J and Williams, Neal M and
                  Blüthgen, Nico and Krasnov, Boris R and Poulin, Robert},
  journaltitle = {Oikos},
  publisher    = {Wiley},
  volume       = {116},
  issue        = {7},
  pages        = {1120--1127},
  date         = {2007-07-01},
  urldate      = {2026-05-19},
  language     = {en},
  doi          = {10.1111/j.0030-1299.2007.15828.x}
}

@ARTICLE{Dorado2011-dw,
  title        = {Rareness and specialization in plant-pollinator networks},
  author       = {Dorado, Jimena and Vázquez, Diego P and Stevani, Erica L and
                  Chacoff, Natacha P},
  journaltitle = {Ecology},
  publisher    = {Wiley},
  volume       = {92},
  issue        = {1},
  pages        = {19--25},
  date         = {2011-01-01},
  urldate      = {2026-05-19},
  language     = {en},
  doi          = {10.1890/10-0794.1}
}

@INPROCEEDINGS{Levy2015-os,
  title      = {Do supervised distributional methods really learn lexical
                inference relations?},
  author     = {Levy, Omer and Remus, Steffen and Biemann, Chris and Dagan, Ido},
  booktitle  = {Proceedings of the 2015 Conference of the North American Chapter
                of the Association for Computational Linguistics: Human Language
                Technologies},
  publisher  = {Association for Computational Linguistics},
  venue      = {Denver, Colorado},
  pages      = {970--976},
  date       = {2015},
  urldate    = {2026-05-19},
  doi        = {10.3115/v1/N15-1098}
}

@INPROCEEDINGS{Kotlerman2009-zo,
  title     = {Directional Distributional Similarity for Lexical Expansion},
  author    = {Kotlerman, Lili and Dagan, Ido and Szpektor, Idan and Geffet,
               Maayan},
  booktitle = {Proceedings of the ACL-IJCNLP 2009 Conference Short Papers},
  pages     = {69--72},
  date      = {2009},
  urldate   = {2026-05-19}
}

@INPROCEEDINGS{Herbelot2013-vj,
  title     = {Measuring semantic content in distributional vectors},
  author    = {Herbelot, Aurélie and Ganesalingam, Mohan},
  booktitle = {Proceedings of the 51st Annual Meeting of the Association for
               Computational Linguistics (Volume 2: Short Papers)},
  pages     = {440--445},
  date      = {2013},
  urldate   = {2026-05-19}
}

@INPROCEEDINGS{Bott2021-pt,
  title      = {More than just frequency? Demasking unsupervised hypernymy
                prediction methods},
  author     = {Bott, Thomas and Schlechtweg, Dominik and Schulte im Walde,
                Sabine},
  booktitle  = {Findings of the Association for Computational Linguistics:
                ACL-IJCNLP 2021},
  publisher  = {Association for Computational Linguistics},
  venue      = {Online},
  pages      = {186--192},
  date       = {2021},
  urldate    = {2026-05-19},
  doi        = {10.18653/v1/2021.findings-acl.16}
}

@INPROCEEDINGS{Geffet2005-fb,
  title      = {The distributional inclusion hypotheses and lexical entailment},
  author     = {Geffet, Maayan and Dagan, Ido},
  booktitle  = {Proceedings of the 43rd Annual Meeting on Association for
                Computational Linguistics - ACL '05},
  publisher  = {Association for Computational Linguistics},
  venue      = {Ann Arbor, Michigan},
  pages      = {107--114},
  date       = {2005},
  urldate    = {2026-05-19},
  doi        = {10.3115/1219840.1219854}
}

@ARTICLE{Bluthgen2006-qv,
  title        = {Measuring specialization in species interaction networks},
  author       = {Blüthgen, Nico and Menzel, Florian and Blüthgen, Nils},
  journaltitle = {BMC Ecol.},
  publisher    = {Springer Nature},
  volume       = {6},
  issue        = {1},
  pages        = {9},
  date         = {2006-08-14},
  language     = {en},
  doi          = {10.1186/1472-6785-6-9}
}

@INPROCEEDINGS{Lee1999-nk,
  title      = {Measures of distributional similarity},
  author     = {Lee, Lillian},
  booktitle  = {Proceedings of the 37th annual meeting of the Association for
                Computational Linguistics on Computational Linguistics -},
  publisher  = {Association for Computational Linguistics},
  venue      = {College Park, Maryland},
  pages      = {25--32},
  date       = {1999},
  language   = {en},
  doi        = {10.3115/1034678.1034693}
}

@techreport{oai:nistep.repo.nii.ac.jp:02000286,
  title      = {An Empirical Analysis of Preprint Lead Time over Peer-Reviewed Articles: Regarding COVID-19/SARS-CoV-2 Research (2020--2025)},
  author     = {OSAKI, Hayami and ITO, Yuko and HAYASHI, Kazuhiro and KOSHIBA, Hitoshi},
  institution  = {National Institute of Science and Technology Policy},
  type       = {NISTEP Discussion Paper},
  number     = {248},
  month      = {Mar},
  year       = {2026},
  language   = {jp},
  doi        = {10.15108/dp248}
}

\clearpage
\begin{appendices}

\section{Baseline numerical reference}
\label{app:baseline-reference}

This appendix reports the numerical baseline underlying the case study sections of the main text. Unless otherwise noted, the baseline specification is
\[
K = 500,\qquad \lambda = 10^{-3},\qquad
\tau = \frac{1}{n_{\min}} = \frac{1}{1756} \approx 5.7\times 10^{-4}.
\]
All quantities in Appendix~\ref{app:baseline-reference} are computed from the full data without rarefaction, except where rarefaction sign-stability summaries are explicitly reported.

\subsection{Within-server values and rankings}

Table~\ref{tab:app-baseline-server} reports the within-server baseline values of \(D_1\), \(D_2\), and \(G\), together with descending ranks (1 = largest value). As in the main text, the purpose is not to force these summaries into a single overall ordering, but to document where the three within-server perspectives diverge.

\begin{table}[h]
\centering
\caption{Within-server baseline values and descending ranks (1 = largest).}
\label{tab:app-baseline-server}
\small
\begin{tabular}{lrrrrrr}
\toprule
Server & \(D_1\) & Rank & \(D_2\) & Rank & \(G\) & Rank \\
\midrule
arXiv & 165.368 & 3 & 125.970 & 3 & 0.272 & 4 \\
bioRxiv & 143.837 & 4 & 114.030 & 4 & 0.232 & 5 \\
ChemRxiv & 59.626 & 6 & 41.527 & 6 & 0.362 & 3 \\
medRxiv & 311.960 & 2 & 271.371 & 1 & 0.139 & 6 \\
SSRN & 322.235 & 1 & 220.738 & 2 & 0.378 & 2 \\
SSRN Lancet & 109.247 & 5 & 58.942 & 5 & 0.617 & 1 \\
\bottomrule
\end{tabular}
\end{table}

For convenience, Table~\ref{tab:app-baseline-server-ranks} records the full descending rank orders, both for the full data and for the rarefaction means.

\begin{table}[h]
\centering
\caption{Descending rank orders for within-server summaries.}
\label{tab:app-baseline-server-ranks}
\small
\begin{tabular}{lp{0.36\textwidth}p{0.36\textwidth}}
\toprule
Quantity & Full data & Rarefaction mean \\
\midrule
$D_1$ & SSRN $>$ medRxiv $>$ arXiv $>$ bioRxiv $>$ SSRN Lancet $>$ ChemRxiv & SSRN $>$ medRxiv $>$ arXiv $>$ bioRxiv $>$ SSRN Lancet $>$ ChemRxiv \\
$D_2$ & medRxiv $>$ SSRN $>$ arXiv $>$ bioRxiv $>$ SSRN Lancet $>$ ChemRxiv & medRxiv $>$ SSRN $>$ arXiv $>$ bioRxiv $>$ SSRN Lancet $>$ ChemRxiv \\
$G$ & SSRN Lancet $>$ SSRN $>$ ChemRxiv $>$ arXiv $>$ bioRxiv $>$ medRxiv & SSRN Lancet $>$ ChemRxiv $>$ SSRN $>$ arXiv $>$ bioRxiv $>$ medRxiv \\
\bottomrule
\end{tabular}
\end{table}

\subsection{Full pairwise values}

Across the 15 unordered pairs at the baseline specification,
\[
D_{\mathrm{JS}} \in [0.209,\, 0.961],\qquad
\Delta_{\mathrm{KL}} \in [-2.435,\, 2.038],\qquad
A_{\tau} \in [-0.450,\, 0.248].
\]
Table~\ref{tab:app-baseline-pairs} reports the full pairwise values used in Sections~5.3--5.6. In addition to the full-data quantities, the last two columns give the rarefaction sign-match proportions for \(\Delta_{\mathrm{KL}}\) and \(A_{\tau}\) over 200 rarefaction replicates.

\begin{table}[t]
\centering
\caption{Baseline pairwise values and rarefaction sign stability.} 
\label{tab:app-baseline-pairs}
\scriptsize
\setlength{\tabcolsep}{4pt}
\begin{tabular}{lrrrrrrrrr}
\toprule
Pair & \(D_{\mathrm{JS}}\) & \(D_{\mathrm{KL}}(p\|q)\) & \(D_{\mathrm{KL}}(q\|p)\) & \(\Delta_{\mathrm{KL}}\) & \(M_{\tau}(p\to q)\) & \(M_{\tau}(q\to p)\) & \(A_{\tau}\) & Sign match \(\Delta_{\mathrm{KL}}\) & Sign match \(A_{\tau}\) \\
\midrule
ChemRxiv--SSRN & 0.781 & 3.954 & 5.991 & -2.036 & 0.557 & 0.801 & -0.245 & 1.000 & 1.000 \\
ChemRxiv--SSRN Lancet & 0.961 & 7.867 & 8.246 & -0.379 & 0.987 & 0.971 & 0.015 & 0.470 & 0.000 \\
ChemRxiv--medRxiv & 0.908 & 7.555 & 6.728 & 0.828 & 0.912 & 0.896 & 0.016 & 1.000 & 0.070 \\
SSRN--SSRN Lancet & 0.501 & 2.642 & 2.430 & 0.212 & 0.517 & 0.350 & 0.168 & 0.690 & 0.995 \\
arXiv--ChemRxiv & 0.789 & 6.092 & 4.054 & 2.038 & 0.712 & 0.464 & 0.248 & 0.995 & 1.000 \\
arXiv--SSRN & 0.469 & 1.968 & 2.737 & -0.769 & 0.295 & 0.484 & -0.188 & 1.000 & 1.000 \\
arXiv--SSRN Lancet & 0.822 & 5.866 & 7.020 & -1.154 & 0.840 & 0.881 & -0.041 & 1.000 & 1.000 \\
arXiv--bioRxiv & 0.824 & 5.538 & 4.991 & 0.547 & 0.879 & 0.797 & 0.082 & 0.945 & 0.990 \\
arXiv--medRxiv & 0.581 & 2.701 & 3.490 & -0.789 & 0.450 & 0.636 & -0.186 & 1.000 & 1.000 \\
bioRxiv--ChemRxiv & 0.643 & 4.017 & 3.397 & 0.620 & 0.397 & 0.401 & -0.004 & 0.230 & 0.070 \\
bioRxiv--SSRN & 0.620 & 2.083 & 4.519 & -2.435 & 0.312 & 0.762 & -0.450 & 1.000 & 1.000 \\
bioRxiv--SSRN Lancet & 0.896 & 6.366 & 7.389 & -1.023 & 0.942 & 0.961 & -0.019 & 1.000 & 1.000 \\
bioRxiv--medRxiv & 0.713 & 3.891 & 4.689 & -0.797 & 0.663 & 0.784 & -0.121 & 0.845 & 1.000 \\
medRxiv--SSRN & 0.209 & 0.646 & 1.091 & -0.445 & 0.121 & 0.247 & -0.126 & 1.000 & 1.000 \\
medRxiv--SSRN Lancet & 0.509 & 2.082 & 3.042 & -0.960 & 0.426 & 0.618 & -0.192 & 1.000 & 0.990 \\
\bottomrule
\end{tabular}
\end{table}

Table~\ref{tab:app-baseline-pair-contrasts} supplements Table~\ref{tab:app-baseline-pairs} with (i) descending ranks for the three pairwise quantities used most directly in the main text and (ii) the within-object contrast quantities \(\Delta D_1\), \(\Delta D_2\), and \(\Delta G\). Ranks are descending (1 = largest value).

\begin{table}[t]
\centering
\caption{Baseline pairwise ranks and within-object contrast quantities.}
\label{tab:app-baseline-pair-contrasts}
\scriptsize
\setlength{\tabcolsep}{4pt}
\begin{tabular}{lrrrrrr}
\toprule
Pair & Rank \(D_{\mathrm{JS}}\) & Rank \(\Delta_{\mathrm{KL}}\) & Rank \(A_{\tau}\) & \(\Delta D_1\) & \(\Delta D_2\) & \(\Delta G\) \\
\midrule
ChemRxiv--SSRN & 7 & 14 & 14 & -262.609 & -179.211 & -0.017 \\
ChemRxiv--SSRN Lancet & 1 & 6 & 5 & -49.622 & -17.415 & -0.255 \\
ChemRxiv--medRxiv & 2 & 2 & 4 & -252.334 & -229.844 & 0.222 \\
SSRN--SSRN Lancet & 13 & 5 & 2 & 212.987 & 161.796 & -0.239 \\
arXiv--ChemRxiv & 6 & 1 & 1 & 105.742 & 84.443 & -0.090 \\
arXiv--SSRN & 14 & 8 & 12 & -156.867 & -94.768 & -0.106 \\
arXiv--SSRN Lancet & 5 & 13 & 8 & 56.120 & 67.028 & -0.345 \\
arXiv--bioRxiv & 4 & 4 & 3 & 21.530 & 11.940 & 0.040 \\
arXiv--medRxiv & 11 & 9 & 11 & -146.592 & -145.401 & 0.133 \\
bioRxiv--ChemRxiv & 9 & 3 & 6 & 84.212 & 72.503 & -0.130 \\
bioRxiv--SSRN & 10 & 15 & 15 & -178.397 & -106.708 & -0.146 \\
bioRxiv--SSRN Lancet & 3 & 12 & 7 & 34.590 & 55.088 & -0.385 \\
bioRxiv--medRxiv & 8 & 10 & 9 & -168.122 & -157.341 & 0.093 \\
medRxiv--SSRN & 15 & 7 & 10 & -10.275 & 50.633 & -0.239 \\
medRxiv--SSRN Lancet & 12 & 11 & 13 & 202.712 & 212.429 & -0.478 \\
\bottomrule
\end{tabular}
\end{table}

For ease of inspection, Table~\ref{tab:app-baseline-pair-ranks} records the full descending rank orders used in the main-text rank comparisons. At the baseline specification, the corresponding Spearman correlations are
\[
\rho(\Delta_{\mathrm{KL}},A_{\tau}) \approx 0.807,
\qquad
\rho(\Delta_{\mathrm{KL}},\Delta G) \approx 0.350,
\qquad
\rho(\Delta_{\mathrm{KL}},\Delta D_1) \approx 0.314,
\qquad
\rho(\Delta_{\mathrm{KL}},\Delta D_2) \approx 0.204.
\]

\begin{table}[p]
\centering
\caption{Descending rank orders for the pairwise quantities used in Section~5.6.}
\label{tab:app-baseline-pair-ranks}
\scriptsize
\setlength{\tabcolsep}{4pt}
\renewcommand{\arraystretch}{0.95}

\textbf{$D_{\mathrm{JS}}$}

\vspace{2mm}

\begin{tabular}{c p{0.33\textwidth} p{0.33\textwidth}}
\toprule
Rank & Full data & Rarefaction mean \\
\midrule
1  & ChemRxiv--SSRN Lancet & ChemRxiv--SSRN Lancet \\
2  & ChemRxiv--medRxiv & ChemRxiv--medRxiv \\
3  & bioRxiv--SSRN Lancet & bioRxiv--SSRN Lancet \\
4  & arXiv--bioRxiv & arXiv--bioRxiv \\
5  & arXiv--SSRN Lancet & arXiv--SSRN Lancet \\
6  & arXiv--ChemRxiv & arXiv--ChemRxiv \\
7  & ChemRxiv--SSRN & ChemRxiv--SSRN \\
8  & bioRxiv--medRxiv & bioRxiv--medRxiv \\
9  & bioRxiv--ChemRxiv & bioRxiv--SSRN \\
10 & bioRxiv--SSRN & bioRxiv--ChemRxiv \\
11 & arXiv--medRxiv & arXiv--medRxiv \\
12 & medRxiv--SSRN Lancet & medRxiv--SSRN Lancet \\
13 & SSRN--SSRN Lancet & SSRN--SSRN Lancet \\
14 & arXiv--SSRN & arXiv--SSRN \\
15 & medRxiv--SSRN & medRxiv--SSRN \\
\bottomrule
\end{tabular}

\vspace{4mm}

\textbf{$\Delta_{\mathrm{KL}}$}

\vspace{2mm}

\begin{tabular}{c p{0.33\textwidth} p{0.33\textwidth}}
\toprule
Rank & Full data & Rarefaction mean \\
\midrule
1  & arXiv--ChemRxiv & ChemRxiv--medRxiv \\
2  & ChemRxiv--medRxiv & arXiv--ChemRxiv \\
3  & bioRxiv--ChemRxiv & arXiv--bioRxiv \\
4  & arXiv--bioRxiv & SSRN--SSRN Lancet \\
5  & SSRN--SSRN Lancet & ChemRxiv--SSRN Lancet \\
6  & ChemRxiv--SSRN Lancet & bioRxiv--ChemRxiv \\
7  & medRxiv--SSRN & bioRxiv--medRxiv \\
8  & arXiv--SSRN & arXiv--medRxiv \\
9  & arXiv--medRxiv & medRxiv--SSRN \\
10 & bioRxiv--medRxiv & bioRxiv--SSRN Lancet \\
11 & medRxiv--SSRN Lancet & arXiv--SSRN \\
12 & bioRxiv--SSRN Lancet & arXiv--SSRN Lancet \\
13 & arXiv--SSRN Lancet & ChemRxiv--SSRN \\
14 & ChemRxiv--SSRN & medRxiv--SSRN Lancet \\
15 & bioRxiv--SSRN & bioRxiv--SSRN \\
\bottomrule
\end{tabular}

\vspace{4mm}

\textbf{$A_{\tau}$}

\vspace{2mm}

\begin{tabular}{c p{0.33\textwidth} p{0.33\textwidth}}
\toprule
Rank & Full data & Rarefaction mean \\
\midrule
1  & arXiv--ChemRxiv & arXiv--ChemRxiv \\
2  & SSRN--SSRN Lancet & SSRN--SSRN Lancet \\
3  & arXiv--bioRxiv & arXiv--bioRxiv \\
4  & ChemRxiv--medRxiv & bioRxiv--ChemRxiv \\
5  & ChemRxiv--SSRN Lancet & ChemRxiv--medRxiv \\
6  & bioRxiv--ChemRxiv & ChemRxiv--SSRN Lancet \\
7  & bioRxiv--SSRN Lancet & bioRxiv--SSRN Lancet \\
8  & arXiv--SSRN Lancet & arXiv--SSRN Lancet \\
9  & bioRxiv--medRxiv & medRxiv--SSRN \\
10 & medRxiv--SSRN & bioRxiv--medRxiv \\
11 & arXiv--medRxiv & arXiv--medRxiv \\
12 & arXiv--SSRN & medRxiv--SSRN Lancet \\
13 & medRxiv--SSRN Lancet & arXiv--SSRN \\
14 & ChemRxiv--SSRN & ChemRxiv--SSRN \\
15 & bioRxiv--SSRN & bioRxiv--SSRN \\
\bottomrule
\end{tabular}
\end{table}

\clearpage

\section{Robustness to smoothing, category resolution, and sample-size standardisation}
\label{app:robustness}

This appendix reports robustness checks for the baseline choice \(K=500\) and \(\lambda=10^{-3}\). The goal is not to identify uniquely correct values of \(K\) or \(\lambda\), but to show how strongly the numerical summaries in Section~5 depend on the representation and smoothing choices.

\subsection{Within-server robustness over the \((K,\lambda)\) grid}

Table~\ref{tab:app-robust-server} reports, for each \((K,\lambda)\), the Spearman rank correlation between the corresponding within-server ordering and the baseline ordering. The full-data ordering of \(D_2\) is completely stable across the grid, whereas \(D_1\) and especially \(G\) show some sensitivity under coarser and/or more heavily smoothed specifications. The rarefaction means show the same broad pattern.

\begin{landscape}
\small
\begin{longtable}{rrcccccc}
\caption{Rank stability of within-server orderings relative to the baseline specification \((K=500,\lambda=10^{-3})\).} \label{tab:app-robust-server}\\
\toprule
\(K\) & \(\lambda\) & \(\rho_{D_1}^{\mathrm{full}}\) & \(\rho_{D_2}^{\mathrm{full}}\) & \(\rho_{G}^{\mathrm{full}}\) & \(\rho_{D_1}^{\mathrm{raref}}\) & \(\rho_{D_2}^{\mathrm{raref}}\) & \(\rho_{G}^{\mathrm{raref}}\) \\
\midrule
\endfirsthead
\toprule
\(K\) & \(\lambda\) & \(\rho_{D_1}^{\mathrm{full}}\) & \(\rho_{D_2}^{\mathrm{full}}\) & \(\rho_{G}^{\mathrm{full}}\) & \(\rho_{D_1}^{\mathrm{raref}}\) & \(\rho_{D_2}^{\mathrm{raref}}\) & \(\rho_{G}^{\mathrm{raref}}\) \\
\midrule
\endhead
300 & $10^{-8}$ & 1.000 & 1.000 & 1.000 & 1.000 & 1.000 & 1.000 \\
300 & $10^{-6}$ & 1.000 & 1.000 & 1.000 & 1.000 & 1.000 & 1.000 \\
300 & $10^{-4}$ & 1.000 & 1.000 & 1.000 & 1.000 & 1.000 & 1.000 \\
300 & $10^{-3}$ & 1.000 & 1.000 & 1.000 & 1.000 & 1.000 & 1.000 \\
300 & $10^{-2}$ & 1.000 & 1.000 & 0.886 & 1.000 & 1.000 & 0.943 \\
300 & $10^{-1}$ & 0.943 & 1.000 & 0.886 & 1.000 & 1.000 & 0.886 \\
500 & $10^{-8}$ & 1.000 & 1.000 & 1.000 & 1.000 & 1.000 & 0.943 \\
500 & $10^{-6}$ & 1.000 & 1.000 & 1.000 & 1.000 & 1.000 & 1.000 \\
500 & $10^{-4}$ & 1.000 & 1.000 & 1.000 & 1.000 & 1.000 & 1.000 \\
500 & $10^{-3}$ & 1.000 & 1.000 & 1.000 & 1.000 & 1.000 & 0.943 \\
500 & $10^{-2}$ & 1.000 & 1.000 & 0.943 & 1.000 & 1.000 & 0.943 \\
500 & $10^{-1}$ & 1.000 & 1.000 & 0.657 & 1.000 & 1.000 & 0.657 \\
800 & $10^{-8}$ & 0.943 & 1.000 & 1.000 & 1.000 & 1.000 & 0.943 \\
800 & $10^{-6}$ & 0.943 & 1.000 & 1.000 & 1.000 & 1.000 & 0.943 \\
800 & $10^{-4}$ & 0.943 & 1.000 & 1.000 & 1.000 & 1.000 & 0.943 \\
800 & $10^{-3}$ & 0.943 & 1.000 & 1.000 & 1.000 & 1.000 & 0.943 \\
800 & $10^{-2}$ & 0.943 & 1.000 & 0.943 & 1.000 & 1.000 & 0.943 \\
800 & $10^{-1}$ & 0.943 & 1.000 & 0.829 & 1.000 & 1.000 & 0.657 \\
1000 & $10^{-8}$ & 0.943 & 1.000 & 1.000 & 1.000 & 1.000 & 0.943 \\
1000 & $10^{-6}$ & 0.943 & 1.000 & 1.000 & 1.000 & 1.000 & 0.943 \\
1000 & $10^{-4}$ & 0.943 & 1.000 & 1.000 & 1.000 & 1.000 & 0.943 \\
1000 & $10^{-3}$ & 0.943 & 1.000 & 1.000 & 1.000 & 1.000 & 0.943 \\
1000 & $10^{-2}$ & 0.943 & 1.000 & 0.943 & 1.000 & 1.000 & 0.943 \\
1000 & $10^{-1}$ & 0.943 & 1.000 & 0.829 & 1.000 & 1.000 & 0.657 \\
1500 & $10^{-8}$ & 0.943 & 1.000 & 1.000 & 0.943 & 1.000 & 0.943 \\
1500 & $10^{-6}$ & 0.943 & 1.000 & 1.000 & 0.943 & 1.000 & 0.943 \\
1500 & $10^{-4}$ & 0.943 & 1.000 & 1.000 & 0.943 & 1.000 & 0.943 \\
1500 & $10^{-3}$ & 0.943 & 1.000 & 1.000 & 0.943 & 1.000 & 0.943 \\
1500 & $10^{-2}$ & 0.943 & 1.000 & 1.000 & 0.943 & 1.000 & 0.943 \\
1500 & $10^{-1}$ & 0.943 & 1.000 & 0.886 & 0.943 & 1.000 & 0.486 \\
\bottomrule
\end{longtable}
\end{landscape}

\subsection{Pairwise robustness over the \((K,\lambda)\) grid}

For pairwise quantities, robustness is evaluated by the Spearman rank correlation between each \((K,\lambda)\) specification and the baseline ordering over the 15 unordered server pairs. Table~\ref{tab:app-robust-pair} reports these values for \(D_{\mathrm{JS}}\), \(\Delta_{\mathrm{KL}}\), and \(A_{\tau}\), both for the full data and for the rarefaction means.

The resulting picture is compact. \(D_{\mathrm{JS}}\) remains highly stable across the grid. \(\Delta_{\mathrm{KL}}\) is most stable under moderate smoothing and deteriorates under very heavy smoothing. The behaviour of \(A_\tau\) is more mixed: it is highly stable across smoothing choices at smaller \(K\), but the full-data ordering becomes less stable at larger \(K\), where the thresholded low-probability diagnostic is more sensitive to the sparsity of the discretized representation. The rarefaction means are more stable, but the full-data values indicate that \(A_\tau\) should be treated as representation-sensitive.

\begin{landscape}
\small
\begin{longtable}{rrcccccc}
\caption{Rank stability of pairwise quantities relative to the baseline specification \((K=500,\lambda=10^{-3})\).} \label{tab:app-robust-pair}\\
\toprule
\(K\) & \(\lambda\) & \(\rho_{D_{\mathrm{JS}}}^{\mathrm{full}}\) & \(\rho_{\Delta_{\mathrm{KL}}}^{\mathrm{full}}\) & \(\rho_{A_{\tau}}^{\mathrm{full}}\) & \(\rho_{D_{\mathrm{JS}}}^{\mathrm{raref}}\) & \(\rho_{\Delta_{\mathrm{KL}}}^{\mathrm{raref}}\) & \(\rho_{A_{\tau}}^{\mathrm{raref}}\) \\
\midrule
\endfirsthead
\toprule
\(K\) & \(\lambda\) & \(\rho_{D_{\mathrm{JS}}}^{\mathrm{full}}\) & \(\rho_{\Delta_{\mathrm{KL}}}^{\mathrm{full}}\) & \(\rho_{A_{\tau}}^{\mathrm{full}}\) & \(\rho_{D_{\mathrm{JS}}}^{\mathrm{raref}}\) & \(\rho_{\Delta_{\mathrm{KL}}}^{\mathrm{raref}}\) & \(\rho_{A_{\tau}}^{\mathrm{raref}}\) \\
\midrule
\endhead
300 & $10^{-8}$ & 0.996 & 0.621 & 0.986 & 0.996 & 0.954 & 0.925 \\
300 & $10^{-6}$ & 0.996 & 0.843 & 0.986 & 0.996 & 0.954 & 0.939 \\
300 & $10^{-4}$ & 0.996 & 0.911 & 0.986 & 0.993 & 0.975 & 0.925 \\
300 & $10^{-3}$ & 0.996 & 0.964 & 0.986 & 0.993 & 0.986 & 0.925 \\
300 & $10^{-2}$ & 0.996 & 0.861 & 0.968 & 0.993 & 0.925 & 0.925 \\
300 & $10^{-1}$ & 0.993 & 0.214 & 0.789 & 0.993 & 0.357 & 0.889 \\
500 & $10^{-8}$ & 1.000 & 0.743 & 1.000 & 0.996 & 0.918 & 0.993 \\
500 & $10^{-6}$ & 1.000 & 0.807 & 1.000 & 1.000 & 0.968 & 0.996 \\
500 & $10^{-4}$ & 1.000 & 0.907 & 1.000 & 1.000 & 0.986 & 0.996 \\
500 & $10^{-3}$ & 1.000 & 1.000 & 1.000 & 1.000 & 1.000 & 1.000 \\
500 & $10^{-2}$ & 1.000 & 0.893 & 0.993 & 0.993 & 0.846 & 0.989 \\
500 & $10^{-1}$ & 0.993 & 0.225 & 0.968 & 0.989 & 0.346 & 0.971 \\
800 & $10^{-8}$ & 0.993 & 0.746 & 0.932 & 0.993 & 0.893 & 0.975 \\
800 & $10^{-6}$ & 0.993 & 0.789 & 0.932 & 0.996 & 0.925 & 0.979 \\
800 & $10^{-4}$ & 0.993 & 0.904 & 0.932 & 0.993 & 0.964 & 0.975 \\
800 & $10^{-3}$ & 0.993 & 0.968 & 0.932 & 0.996 & 0.982 & 0.979 \\
800 & $10^{-2}$ & 0.996 & 0.882 & 0.921 & 0.989 & 0.804 & 0.968 \\
800 & $10^{-1}$ & 0.989 & 0.225 & 0.932 & 0.989 & 0.243 & 0.957 \\
1000 & $10^{-8}$ & 0.989 & 0.764 & 0.639 & 0.993 & 0.921 & 0.971 \\
1000 & $10^{-6}$ & 0.989 & 0.807 & 0.639 & 0.993 & 0.943 & 0.982 \\
1000 & $10^{-4}$ & 0.989 & 0.861 & 0.639 & 0.993 & 0.982 & 0.971 \\
1000 & $10^{-3}$ & 0.989 & 0.961 & 0.639 & 0.993 & 0.943 & 0.971 \\
1000 & $10^{-2}$ & 0.993 & 0.793 & 0.639 & 0.993 & 0.696 & 0.971 \\
1000 & $10^{-1}$ & 0.989 & 0.196 & 0.921 & 0.989 & 0.207 & 0.896 \\
1500 & $10^{-8}$ & 0.979 & 0.754 & 0.464 & 0.989 & 0.900 & 0.918 \\
1500 & $10^{-6}$ & 0.979 & 0.771 & 0.464 & 0.989 & 0.936 & 0.918 \\
1500 & $10^{-4}$ & 0.979 & 0.871 & 0.464 & 0.989 & 0.975 & 0.918 \\
1500 & $10^{-3}$ & 0.979 & 0.929 & 0.464 & 0.989 & 0.875 & 0.918 \\
1500 & $10^{-2}$ & 0.975 & 0.789 & 0.461 & 0.989 & 0.546 & 0.918 \\
1500 & $10^{-1}$ & 0.968 & 0.143 & 0.625 & 0.979 & 0.164 & 0.861 \\
\bottomrule
\end{longtable}
\end{landscape}

\clearpage
\section{Sensitivity of \(A_{\tau}\) to the threshold choice}
\label{app:tau-sensitivity}

This appendix reports the threshold sensitivity checks referred to in Section~5.4. The baseline analysis uses
\[
\tau = \frac{1}{n_{\min}} = \frac{1}{1756} \approx 5.7\times 10^{-4},
\]
and the sweep below examines the local grid
\[
\tau \in \left\{\tfrac{1}{4n_{\min}}, \tfrac{1}{2n_{\min}}, \tfrac{1}{n_{\min}}, \tfrac{2}{n_{\min}}, \tfrac{4}{n_{\min}}\right\}.
\]

Table~\ref{tab:app-tau-summary} reports the full-data and rarefaction summaries. Over the range
\[
\tau \in \left\{\tfrac{1}{4n_{\min}}, \tfrac{1}{2n_{\min}}, \tfrac{1}{n_{\min}}\right\},
\]
the rank association between \(\Delta_{\mathrm{KL}}\) and \(A_{\tau}\) remains close to \(0.80\), both in the full data and on average under rarefaction. The association weakens at \(2/n_{\min}\) and collapses by \(4/n_{\min}\). The sign-match rates show the same pattern.

\begin{table}[h]
\centering
\caption{Threshold-sensitivity summary for \(A_{\tau}\) at the baseline specification.}
\label{tab:app-tau-summary}
\scriptsize
\begin{tabular}{lrrrrrrr}
\toprule
Threshold & \(\tau\) & \(\rho(\Delta_{\mathrm{KL}},A_{\tau})\) full & Sign match full & \(\rho\) raref mean & \(\rho\) raref sd & Sign match raref mean & Sign match raref sd \\
\midrule
$0.25/n_{\min}$ & 0.000142 & 0.804 & 0.933 & 0.796 & 0.054 & 0.829 & 0.060 \\
$0.5/n_{\min}$ & 0.000285 & 0.800 & 0.867 & 0.796 & 0.054 & 0.829 & 0.060 \\
$1/n_{\min}$ & 0.000569 & 0.807 & 0.867 & 0.796 & 0.054 & 0.829 & 0.060 \\
$2/n_{\min}$ & 0.001139 & 0.671 & 0.733 & 0.744 & 0.060 & 0.817 & 0.065 \\
$4/n_{\min}$ & 0.002278 & 0.143 & 0.533 & -0.047 & 0.123 & 0.452 & 0.092 \\
\bottomrule
\end{tabular}
\end{table}

Table~\ref{tab:app-tau-pairs} shows the same sweep for the two illustrative pairs used in Section~5.5. The higher-overlap illustration (medRxiv--SSRN) is stable up to \(2/n_{\min}\), whereas the weak-overlap cautionary pair (ChemRxiv--medRxiv) remains close to zero and changes sign across nearby thresholds. This is precisely the pattern the main text is designed to diagnose: a pair with a substantial and robust low-probability asymmetry is a good candidate for substantive interpretation, whereas a pair whose \(A_{\tau}\) remains near zero and threshold-fragile should be read cautiously even if \(\Delta_{\mathrm{KL}}\) itself is numerically large.

\begin{landscape}
\small
\begin{longtable}{llrrrrrr}
\caption{Threshold sweep for the two illustrative pairs.} \label{tab:app-tau-pairs}\\
\toprule
Pair & Threshold & \(A_{\tau}\) full & \(M_{\tau}(p\to q)\) full & \(M_{\tau}(q\to p)\) full & \(A_{\tau}\) raref mean & \(A_{\tau}\) raref sd & Sign match raref \\
\midrule
\endfirsthead
\toprule
Pair & Threshold & \(A_{\tau}\) full & \(M_{\tau}(p\to q)\) full & \(M_{\tau}(q\to p)\) full & \(A_{\tau}\) raref mean & \(A_{\tau}\) raref sd & Sign match raref \\
\midrule
\endhead
ChemRxiv--medRxiv & $0.25/n_{\min}$ & -0.081 & 0.815 & 0.896 & -0.030 & 0.022 & 0.085 \\
ChemRxiv--medRxiv & $0.5/n_{\min}$ & -0.070 & 0.825 & 0.896 & -0.030 & 0.022 & 0.085 \\
ChemRxiv--medRxiv & $1/n_{\min}$ & 0.016 & 0.912 & 0.896 & -0.030 & 0.022 & 0.085 \\
ChemRxiv--medRxiv & $2/n_{\min}$ & -0.003 & 0.954 & 0.957 & -0.024 & 0.014 & 0.035 \\
ChemRxiv--medRxiv & $4/n_{\min}$ & -0.001 & 0.984 & 0.985 & -0.002 & 0.004 & 0.375 \\
medRxiv--SSRN & $0.25/n_{\min}$ & -0.155 & 0.016 & 0.171 & -0.149 & 0.022 & 1.000 \\
medRxiv--SSRN & $0.5/n_{\min}$ & -0.144 & 0.053 & 0.197 & -0.149 & 0.022 & 1.000 \\
medRxiv--SSRN & $1/n_{\min}$ & -0.126 & 0.121 & 0.247 & -0.149 & 0.022 & 1.000 \\
medRxiv--SSRN & $2/n_{\min}$ & -0.082 & 0.249 & 0.330 & -0.120 & 0.021 & 1.000 \\
medRxiv--SSRN & $4/n_{\min}$ & 0.095 & 0.606 & 0.511 & 0.062 & 0.025 & 0.000 \\
\bottomrule
\end{longtable}
\end{landscape}

\end{appendices}

\end{document}